\documentclass[preprint,floatfix,amsmath,amssymb,aps,onecolumn,nolongbibliography]{revtex4-2}

\usepackage[english]{babel}

\usepackage[hidelinks=true]{hyperref}
\hypersetup{
    colorlinks,
    linkcolor={blue!80!black},
    citecolor={blue!80!black},
    urlcolor={blue!80!black}
}

\usepackage{setspace}
\usepackage{caption}
\usepackage{siunitx}
\usepackage{subcaption}
\usepackage{graphicx}
\usepackage{dcolumn}
\usepackage{bm}
\usepackage{gensymb}
\usepackage{lipsum}
\usepackage{mathrsfs}  
\usepackage[thinc]{esdiff}  

\usepackage{xcolor}
\pagecolor{white}

\newcommand{\ie}{\textit{i.e.}}
\newcommand{\eg}{\textit{e.g.}}

\newcommand{\hi}{hydrodynamic interactions }

\begin{document}

\setlength{\unitlength}{1cm}

\title{Hydrodynamic diffusion in apolar active suspensions}

\author{Zhouyang Ge$^{1,2}$}\email{zhoge@mech.kth.se}
\author{Gwynn J.~Elfring$^1$}\email{gelfring@mech.ubc.ca}

\affiliation{$^1$Department of Mechanical Engineering and Institute of Applied Mathematics, University of British Columbia, Vancouver V6T 1Z4, BC, Canada}
\affiliation{$^2$FLOW, Department of Engineering Mechanics, KTH Royal Institute of Technology, 100 44 Stockholm, Sweden}

\begin{abstract} 

Active suspensions encompass a wide range of complex fluids containing microscale energy-injecting particles, such as cells, bacteria or artificially powered active colloids. 
Because they are intrinsically non-equilibrium, active suspensions can display a number of fascinating phenomena, including turbulent-like large-scale coherent motion and enhanced diffusion.
Here, using a recently developed active Fast Stokesian Dynamics method, we present a detailed numerical study on the hydrodynamic diffusion in apolar active suspensions.
Specifically, we simulate suspensions of active but non-self-propelling spherical squirmers, of either puller- or pusher-type, at volume fractions from 0.5\% to 55\%. 
Our results show little difference between pullers and pushers in their instantaneous and long-time dynamics, where the translational dynamics vary non-monotonically with the volume fraction, with a peak diffusivity at around 10\% to 20\%, in stark contrast to suspensions of self-propelling particles.
On the other hand, the rotational dynamics tend to increase with the volume fraction as is the case for self-propelling particles.
To explain these dynamics, we provide detailed scaling and statistical analyses based on the activity-induced \hi and the observed microstructural correlations, which display a weak local order. 
Overall, these results elucidate and highlight the different effects of particle activity on the collective dynamics and transport phenomena in active fluids.

\end{abstract}

\date{\today}
\maketitle
\newpage
\tableofcontents
\newpage

\section{Introduction}

Flowing suspensions of non-Brownian particles often display a stochastic, diffusive-like motion known as \emph{hydrodynamic diffusion} \citep{davis1996hydrodynamic}. 
Common examples include sheared suspensions, where particles driven by an externally imposed velocity gradient undergo chaotic and anisotropic displacements \citep{drazer_koplik_khusid_acrivos_2002, Pine_Nature_2005}, and sedimentation, where long-range velocity correlations result in tortuous and loopy falling paths \citep{guazzelli2011fluctuations}. 
In both cases, the apparent random motions arise not from thermal fluctuations but rather the \emph{flow-induced multiparticle interactions}, determined by the relative positions and orientations of many particles, leading to complex dynamics.

The dynamics may be more complex if the suspending particles are \emph{active}, expending and dissipating energy at the microscale \citep{ramaswamy2017active}. 
In a seminal experiment, \citet{WL2000} studied particle diffusion in a freely suspended soap film containing \emph{E.~coli} bacteria. 
Adding trace amounts of micron-sized polystyrene spheres, they observed short-time superdiffusive and long-time diffusive dynamics of the tracers, with an effective diffusion coefficient 2--3 orders of magnitude higher than the background Brownian motion. 
This enhanced diffusion was attributed to transient formations of swirls and jets, or \emph{coherent structures}, that increase in size and duration with the bacterial concentration.
Many subsequent studies have investigated this mechanism, providing more detailed understanding of particle transport in active suspensions.
For example, collective dynamics of microswimmers were reported in a number of experiments \citep{Dombrowski2004, Sokolov2007, zhang2010collective, wensink2012meso, Lushi2014, li2019data} and simulations \citep{Hernandez-Ortiz2005, Saintillan_Shelley2007, Saintillan_Shelley2008, ishikawa2008coherent, Zottl2014}, suggesting that large-scale coherent motion could emerge purely from \hi \citep{Simha_Ramaswamy2002, Baskaran2009, koch_subramanian_review2011}.
On the other hand, even in dilute suspensions where swimmers' dynamics are approximately uncorrelated, there is an enhanced diffusion proportional to the product of the swimmers' density and their mean speed due to the advective flow they generate \citep{Mino2011, Lin2011, Pshkin2013, Jepson2013, morozov2014enhanced}.
Besides bacteria which swim by pushing the fluid behind, the diffusion of passive particles were also measured in puller-type algal suspensions \citep{Leptos2009, Kurtuldu2011, Yang2016, von2021diffusive}.
An interesting observation in those systems is that the statistics of the particle displacements, though Gaussian at long times, are non-Gaussian transiently \citep{thiffeault2015distribution, ortlieb2019statistics}.

Given the importance of hydrodynamic interactions, many theoretical or numerical models have been proposed to elucidate the rich hydrodynamic effects on particle diffusion in active suspensions.
Graham and coworkers \citep{Hernandez-Ortiz2005, Underhill2008} developed a minimal model, where self-propelled particles were treated as rigid dumbbells exerting force dipoles on the fluid.
The dumbbells were made of point particles and interacted mainly via the fluid flow they generated.
Their simulations showed collective dynamics similar to experimental observations and highlighted some differences between ``pushers" and ``pullers".
\citet{Saintillan_Shelley2007, Saintillan_Shelley2008} used a slender-body model and a kinetic theory to analyze the dynamics of self-propelling rods.
Their results showed that, not only were aligned suspensions hydrodynamically unstable as predicted by an earlier theory  \citep{Simha_Ramaswamy2002}, but there was also an instability in isotropic suspensions of pushers due to stress fluctuations.
Ishikawa, Pedley, and coworkers \citep{ishikawa2007diffusion, Ishikawa2010diffusion, kogure2023flow} modified the Stokesian Dynamics method \citep{sd1988} to simulate the diffusion of finite-size or Lagrangian tracer particles in suspensions of ``squirmers" \citep{Pedley2016}.
Their method provided an accurate description of the near-field hydrodynamic interactions, allowing them to consider more concentrated suspensions.
Finally, many other methods have been proposed or adapted to study transport phenomena in active matter, such as the lattice-Boltzmann method \citep{Llopis2006, Stenhammar2017}, multiparticle collision dynamics \citep{Zottl2014}, dissipative particle dynamics \citep{Chen2016dpd}, and force-coupling method \citep{Delmotte2018}.

Despite the significant development and progress so far, a complete understanding of the underlying \hi between active and passive particles has still not been established.
For one, most previous studies were focused on suspensions of self-propelling particles, thus excluding active but \emph{individually immotile} particles such as melanocytes in the skin \citep{Simha_Ramaswamy2002, Baskaran2009}, microtubule bundles powered by motor proteins \citep{Woodhouse2012, Sanchez2012}, or cells in crowded environments \citep{hallatschek2023}. 
Without separating these two aspects, one cannot delineate how much the observed collective dynamics is due to self-propulsion versus fluid-mediated hydrodynamic interactions.
Furthermore, the passive particles employed earlier were often considered to be tracers that were merely advected by the flow produced by active swimmers.
However, at high enough concentrations, the particles may develop certain \emph{spatial correlations} \citep{Saintillan_Shelley2007, li2019data}, where the dynamics of active particles can be affected by their passive counterpart through both hydrodynamic and excluded-volume interactions.
This is particularly relevant for characterizing realistic active systems, biological or synthetic, since they almost always contain some passive components in the form of dead cells or defects \citep{Jepson2013}.

Motivated by the aforementioned observations, we perform large-scale numerical simulations to study the hydrodynamic diffusion in \emph{apolar} active suspensions.
Specifically, we simulate suspensions of \emph{shakers}, spherical squirmers that are active but not self-propelling, at volume fractions from 0.5\% to 55\%.
The numerical method we use, an ``active version" of the Fast Stokesian Dynamics \citep{fiore2019fast}, accounts for both far-field \hi and near-field lubrication, as well as excluded-volume interactions amongst the particles \citep{elfring2022active}.
Our approach is similar to that of Ishikawa and Pedley, though the previous work did not consider shaker-type squirmers and the number of particles was often quite small (usually around 100).
Here, extensive simulations of more than 1000 shakers show that the instantaneous and long-time translational dynamics are nearly identical between pullers and pushers, which vary \emph{non-monotonically} with the volume fraction, in contrast to suspensions of self-propelling particles.
The rotational dynamics tend to increase with the volume fraction as for self-propelling particles, though the detailed scalings are different.
Remarkably, our results can be well described by the model of \citet{WL2000}, which we show by providing detailed scaling and statistical analyses.
The observed scalings are supported by spatial correlations in the suspension microstructure, which display a weak local alignment due purely to hydrodynamic interactions. 

The paper is organized as follows.
In Section \ref{sec:method}, we describe the mathematical formulation of the squirmer model (\ref{sec:squirmer}) and the governing equations of the active Fast Stokesian Dynamics method (\ref{sec:fsd}), followed by the solver verification (\ref{sec:verifications}) and a brief discussion of pair interactions (\ref{subsec:pair}).
In Section \ref{sec:results}, we detail the numerical setup (\ref{sec:setup}) and present the full simulation results, beginning with the instantaneous speeds and their distributions (\ref{sec:short}), continuing to the hydrodynamic diffusion in both translational and rotational motion (\ref{sec:long}), and ending with an analysis of spatial correlations in the suspension microstructure (\ref{sec:correlations}).
The discussion is focused on the effect of particle volume fraction, though we have also simulated binary suspensions at varying fractions of passive particles (see Appendix \ref{sec:fp}). 
Finally, we conclude in Section \ref{sec:conclusion}.

\section{Models and methods}
\label{sec:method}

\subsection{The squirmer model}
\label{sec:squirmer}

The squirmer model, originally proposed by \citet{lighthill1952} and later extended by \citet{blake1971}, has been widely employed to study the swimming mechanisms and collective dynamics of microorganisms at low Reynolds numbers \citep{Pedley2016}.
Mathematically, the tangential slip velocity on the surface of a spherical squirmer may be expressed as \citep{elfring2022active}
\begin{equation} \label{eq:squirmer}
  {\bm u}_s ({\bm p} \cdot \hat{\bm r}) = \sum_{n=1}^\infty \frac{2}{n(n+1)} P_n^\prime({\bm p} \cdot \hat{\bm r})
  \big [ B_n({\bm I}- \hat{\bm r}\hat{\bm r}) \cdot {\bm p} + C_n {\bm p} \times \hat{\bm r} \big ],
\end{equation}
where ${\bm p}$ is a unit vector defining the axis of a squirmer,
$\hat{\bm r}$ is the unit normal vector on the surface, 
$P_n$ is the Legendre polynomial of degree $n$, and
$P_n^\prime(x)=\textrm{d} P_n(x)/\textrm{d} x$.
Note that, the slip velocity is axisymmetric as ${\bm u}_s$ is a function of ${\bm p}\cdot \hat{\bm r}$ only.

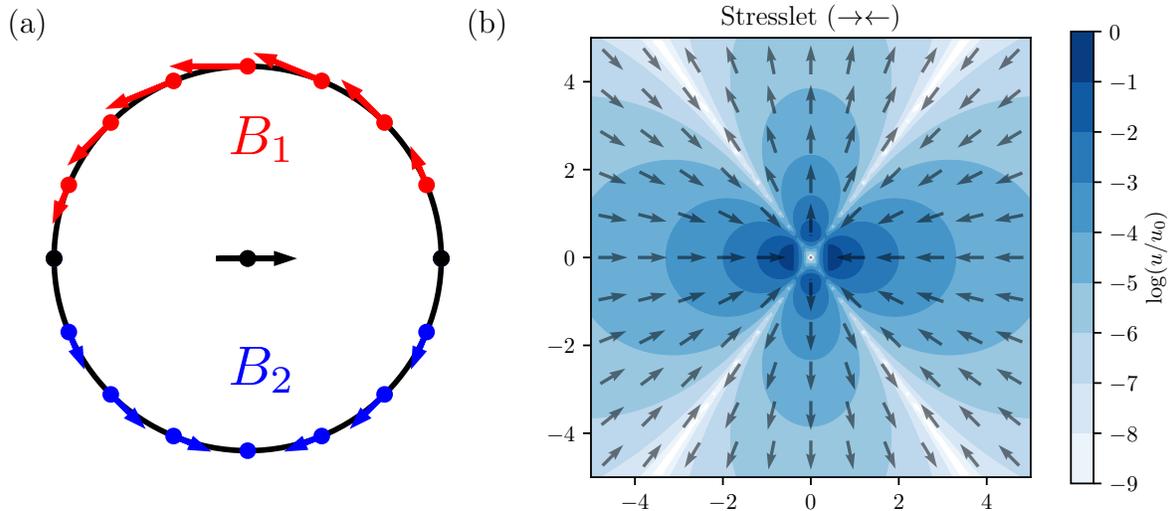
\begin{figure*}[t]
  \centering 
  \includegraphics[height=7.2cm]{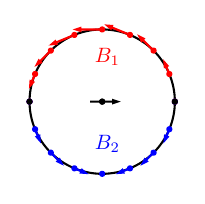} 
  \includegraphics[height=7.2cm]{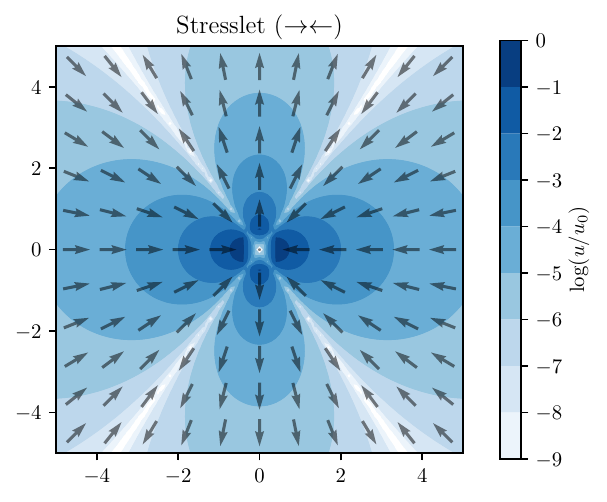} 
  \begin{picture}(0,0)
    \put(-15.8,6.6){(a)} \put(-9.7,6.6){(b)}
  \end{picture} 
  \caption{(a) Slip velocity on a squirmer due to either $B_1$ (red) or $B_2$ (blue).
                (b) The far-field stresslet flow due to $B_2$ (arrows indicate direction while color the magnitude).
                In both cases, the squirmer is oriented horizontally to the right.}
  \label{fig:squirmer}
\end{figure*}

From Eq.~\eqref{eq:squirmer} we can observe two modes of motion: the $B_n$'s are the so-called squirming modes, which gives rise to the longitudinal velocities about ${\bm p}$ (such as the self-propulsion and self-straining); and the $C_n$'s are associated with azimuthal slip velocities about ${\bm p}$ (such as the self-spin).
In the literature, it is customary to neglect all spinning modes while only keep the two leading-order squirming modes \citep{ishikawa2006hydrodynamic, elfring2022active}.
Figure \ref{fig:squirmer} shows the local slip velocity due to $B_1$ and $B_2$, respectively, as well as the far-field stresslet flow due to $B_2$ (see the supplemental information of \citet{mathijssen2015tracer} for the far-field flow patterns of a few higher-order moments).
Note the asymmetry of the flow intensity in the axial and perpendicular directions due to mass conservation in three dimensions (in two dimensions, the axial and perpendicular directions are symmetric, \ie, the velocity is zero along 45 degrees).
Finally, as we will describe in the next section, the prescribed slip velocity in our simulations is simply ${\bm u}_s = {\bm U}_s + {\bm E}_s \cdot \hat{\bm r}$, where ${\bm U}_s$ is the self-propulsion velocity and ${\bm E}_s$ is the self-strain-rate tensor.
Unlike the illustration in Figure \ref{fig:squirmer}(a), this ${\bm u}_s$ is not purely tangential.
However, since the leading order \hi due to activity arise from ${\bm E}_s$, our description can be considered as a minimal model that captures the \hi between squirmers.

\subsection{Active Fast Stokesian Dynamics}
\label{sec:fsd}

The dynamics of a suspension of squirmers can be formulated according to the Active Stokesian Dynamics framework \citep{elfring2022active}, extended from the original Stokesian Dynamics (SD) method \citep{sd1988}.
Assuming small, inertialess particles swimming in a viscous fluid at low Reynolds numbers,
the external forces and torques on any particle must be balanced by their hydrodynamic counterparts,
which are linearly related to the velocity moments on the particle through a hydrodynamic resistance tensor.
In the absence of external flow and Brownian motion, this leads to
\begin{equation} \label{eq:active-sd}
    {\bm U}  = {\bm U}_s + {\bm R}_\text{FU}^{-1} \cdot ({\bm F}_{ext} - {\bm R}_\text{FE}:{\bm E}_s),
\end{equation}
where 
${\bm U} \equiv (\bm U_1 \ \bm U_2 \ ... \ \bm U_N)^T$ denotes the array of velocities for $N$ particles (the same applies to $\bm U_s$, $\bm E_s$, and $\bm F_{ext}$),
and ${\bm R}_\text{FU}$ and ${\bm R}_\text{FE}$ are the resistance tensors coupling the force and velocity moments of \emph{all} particles.
For example, if particle $i$ is active, both $\bm U_{s,i}$ and $\bm E_{s,i}$ can be nonzero ($\bm U_{s,i}=\bm0$ for a shaker); otherwise, $\bm U_{s,i}=\bm E_{s,i} = \bm0$.
In this compact notation, $\bm U_i$ includes both the linear and angular velocities of particle $i$; similarly, $\bm F_{\textrm{ext},i}$ includes both the external force and torque acted on particle $i$.
Finally, because ${\bm R}_\text{FU}$ and ${\bm R}_\text{FE}$ only depend on the positions and orientations of the particles,
the velocities can be computed quasi-statically to evolve the suspension.

Numerically, we solve a modified form of Eq.~\eqref{eq:active-sd} using the Fast Stokesian Dynamics (FSD) method \citep{fiore2019fast}; here, it is considered fast because the computational cost scales linearly with the number of particles.
As in the conventional SD, FSD decomposes the hydrodynamic resistance into a far-field, many-body interaction and a near-field, pairwise contribution.
Specifically, the far-field interaction is obtained by inverting a truncated multipole expansion of the Stokes flow induced by all particles,  
whereas the near-field interaction is the lubrication between two particles minus the duplicated parts in the far-field term.
This leads to the following expressions of the resistance tensors 
\begin{equation} \label{eq:decomp}
  {\bm R}_\text{FU}= \mathcal{B}^T (\mathcal{M}^\text{ff})^{-1} \mathcal{B} + {\bm R}_\text{FU}^\text{nf}, \quad
  {\bm R}_\text{FE}= \mathcal{B}^T (\mathcal{M}^\text{ff})^{-1} \mathcal{C} + {\bm R}_\text{FE}^\text{nf},
\end{equation}
where 
${\bm R}^\text{nf}_\text{FU}$ and ${\bm R}^\text{nf}_\text{FE}$ are the near-field resistance tensors,
and $\mathcal{M}^\text{ff}$ is the mobility tensor relating far-field force moments,
$\mathcal{F}^\text{ff} \equiv ({\bm F}^\text{ff} \ {\bm S}^\text{ff})^T$, 
to the velocity moments,
$\mathcal{U} \equiv ({\bm U} \ {\bm E})^T$, 
through mapping tensors $\mathcal{B}$ and $\mathcal{C}$, 
\begin{equation} \label{eq:maps}
  \mathcal{B} {\bm U}  + \mathcal{C} {\bm E} = \mathcal{U}, \quad
  \mathcal{B}^T \mathcal{F}^\text{ff} = {\bm F}^\text{ff} .
\end{equation}
After some algebra, Eq.~\eqref{eq:active-sd} can then be reformulated as
\begin{equation} \label{eq:afsd}
\begin{bmatrix}
  \mathcal{M}^\text{ff} & \mathcal{B} \\
  \mathcal{B}^T & - \bm{R}_\text{FU}^\text{nf}
\end{bmatrix}
\cdot
\begin{bmatrix} 
  \mathcal{F}^\text{ff} \\ 
  \bm{U}_*
\end{bmatrix}
=
\begin{bmatrix}
  -\mathcal{C} \bm{E}_s\\ 
  \bm{F}_*
\end{bmatrix},
\end{equation}
with the short-hand notation
\begin{equation}  \label{eq:afsd1}
  \bm{U}_* =  \bm{U} - \bm{U}_s , \quad
  \bm{F}_* = - {\bm F}_{ext} + \bm{R}_\text{FE}^\text{nf}: \bm{E}_s .
\end{equation}
This can be verified by substituting Eqs.~(\ref{eq:decomp}, \ref{eq:maps}, \ref{eq:afsd1}) into Eq.~\eqref{eq:afsd} to recover Eq.~\eqref{eq:active-sd}.

Eqs.~(\ref{eq:afsd}, \ref{eq:afsd1}) are the governing equations of our active FSD method.
They can be solved efficiently using fast iterative methods (\eg, Krylov subspace methods) implemented on modern graphical processing units \cite{fiore2019fast}.
For Krylov subspace methods to converge quickly, it is important to precondition the linear system \cite{Benzi_Golub_Liesen_2005};
here, we can use exactly the same preconditioner as in the original FSD since the matrix on the left-hand side of Eq.~\eqref{eq:afsd} does not depend on the particle activity.
Once the solution ${\bm U}$ is obtained, the particle positions and orientations are integrated forward in time using the standard 
second-order Runge-Kutta method;
see \footnote{The source code of our implementation of the active Fast Stokesian Dynamics method is available at \url{https://github.com/GeZhouyang/FSD}.} and \citet{fiore2019fast} for more details of the numerical methods.

\subsection{Solver verification}
\label{sec:verifications}

The original FSD solver has already been verified in colloidal suspensions by \citet{fiore2019fast}, and in noncolloidal dense suspensions by \citet{Ge2022}.
In the following, we present two additional verifications involving a pair of particles where exact or approximate analytical solutions are available.

\begin{figure*}[t]
  \centering
  \includegraphics[height=5.2cm]{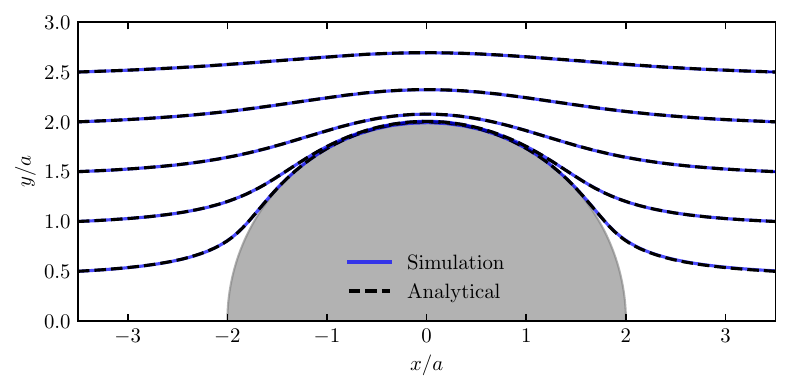}
  \includegraphics[height=5.2cm]{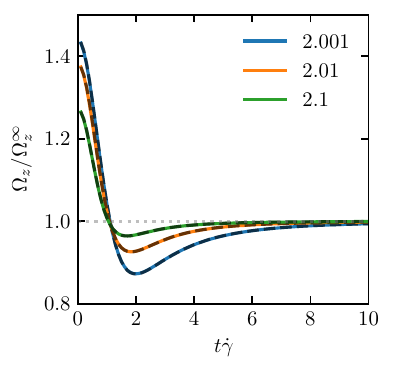}
  \begin{picture}(0,0)
    \put(-16.4,4.7){(a)} \put(-5.7,4.7){(b)}
  \end{picture} 
  \caption{(a) Relative trajectories of two passive spheres in the $xy$ plane and
               (b) their angular velocity $\Omega_z$ (right) in simple shear flow,
               where $a$ is the particle radius, $\dot \gamma$ the shear rate, 
               and $\Omega_z^\infty=\dot\gamma/2$ the angular velocity of the flow.
               The different curves in (b) correspond to the three initial conditions, $y_{min}/a=2.001$, 2.01, and 2.1.
               The analytical solutions (dashed lines) in both (a) and (b) are due to \citet{batchelor1972hydrodynamic}.}
  \label{fig:BG-traj}
\end{figure*}
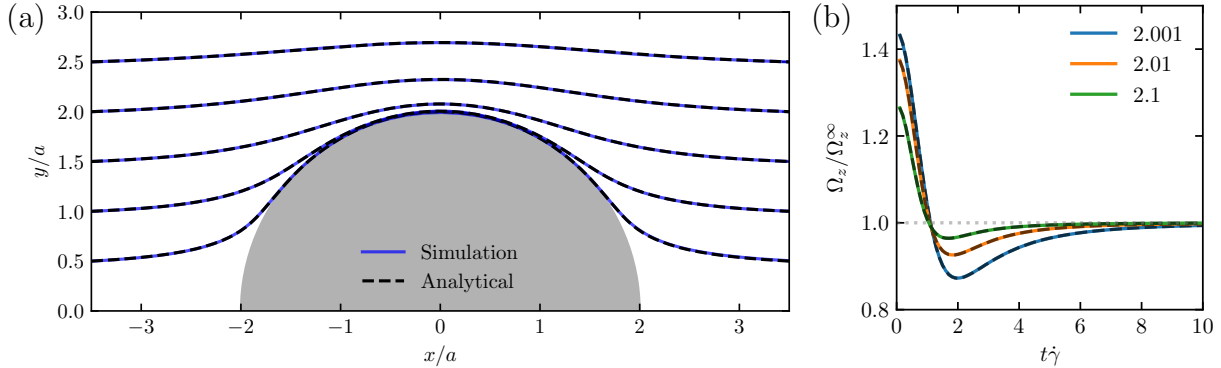

First, we consider the relative motion of two identical passive particles in simple shear flow, for which an analytical solution was provided by \citet{batchelor1972hydrodynamic} (BG).
Figure \ref{fig:BG-traj} shows the simulated trajectories from different initial positions in comparison to the BG solution.
Here, the trajectories are fore-aft symmetric in the absence of non-hydrodynamic interactions or particle roughness.
The interparticle gaps are minimal at $x=0$, when the two particles are aligned in the velocity gradient direction.
The excellent agreement between the simulations and the analytical solutions confirms our numerical calculations of hydrodynamic interactions.
Furthermore, the angular velocities of the particles, though not affecting the translational motion of passive spheres, are also verified. 
This is important for the simulation of active particles as their dynamics depend on the orientations.

Next, we consider the motion of a passive particle next to a shaker without any background flow.
This is the most basic configuration from which the dynamics of a pair of active particles can be constructed, thanks to the linearity of the Stokes equation \citep{ishikawa2006hydrodynamic}. 
In the purely hydrodynamic limit, \ie, no other interactions between the particles, the passive particle is simply advected by the flow generated by the shaker.
Furthermore, if the distance between the two particles is relatively large, we may neglect the disturbance to the original flow field caused by the passive particle, because the disturbance flow decays faster with distance.
Under this \emph{far-field approximation}, the velocity of a force/torque-free passive particle next to a shaker can be obtained by the Fax\'en's law,
\begin{equation} \label{eq:U_p}
  {\bm U}_{p} = {\bm u} +\frac{a^2}{6} \nabla^2 {\bm u}, \quad
  {\bm \Omega}_{p} = \frac{1}{2} \nabla \times {\bm u},
\end{equation}
where ${\bm U}_{p}$ and ${\bm \Omega}_{p}$ are, respectively, the translational and angular velocities of the passive particle, and ${\bm u}$ denotes the flow field generated by an isolated shaker \citep{ishikawa2006hydrodynamic},
\begin{equation} \label{eq:u_sol}
  {\bm u}({\bm p}, {\bm r}) = 
  B_2 \bigg [ 
  \bigg(\frac{a^4}{r^4} - \frac{a^2}{r^2}\bigg)\bigg(\frac{3}{2} \big({\bm p} \cdot \hat{\bm r} \big)^2 -\frac{1}{2}\bigg) \hat{\bm r}
  + \frac{a^4}{r^4} \big({\bm p} \cdot \hat{\bm r} \big) \big({\bm p} \cdot \hat{\bm r} \hat{\bm r} - {\bm p} \big) 
  \bigg ].
\end{equation}
In the above, ${\bm r}=r\hat{\bm r}$ is the distance vector from the center of the shaker to a point in the fluid.
We can see that $|{\bm u}| \sim 1/r^2$ to the leading order and ${\bm u}({\bm p}, {\bm r}) = {\bm u}(-{\bm p}, {\bm r})$, both of which are expected for the dipolar flow generated by a shaker.
Consequently, the hydrodynamically induced velocities follow the scalings $|{\bm U}_{p}| \sim 1/r^2$ and $|{\bm \Omega}_{p}| \sim 1/r^3$ to the leading order. 
One thus expects weaker rotations than translations in dilute suspensions of shakers.

\begin{figure*}[t]
  \centering
  \includegraphics[height=7.5cm]{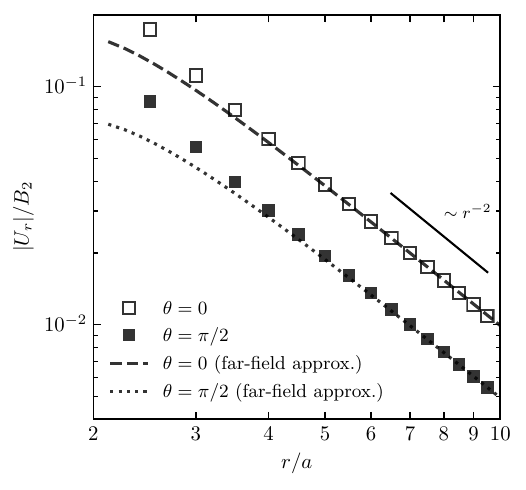} 
  \includegraphics[height=7.5cm]{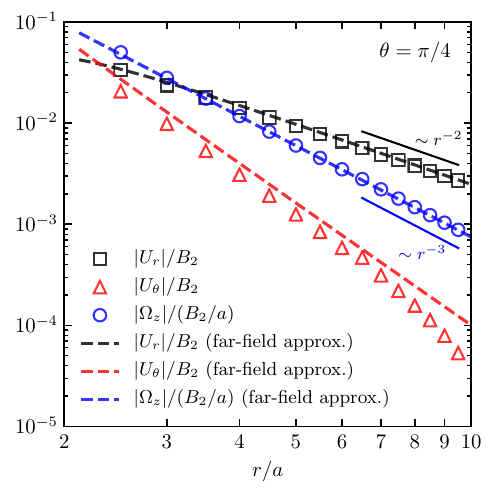} 
  \begin{picture}(0,0)
  \end{picture} 
  \caption{Velocity of a passive particle as a function of distance ($r/a$) next to a pulling shaker, 
                in comparison to the far-field approximation, c.f.~Eqs.~(\ref{eq:U_p}, \ref{eq:u_sol}).
                Here, the passive particle is at angle $\theta$ from the shaker, 
                where $\cos \theta \equiv {\bm p}\cdot \hat{\bm r}$; see the main text for details.}
  \label{fig:Ishikawa}
\end{figure*}

Figure \ref{fig:Ishikawa} shows the velocity of the passive particle as a function of distance ($r/a$) at various angles ($\theta$) away from a pulling shaker.
Here, the translational velocity is decomposed into a radial ($U_r$) and an azimuthal ($U_\theta$) component, and the angular velocity is only nonzero in the perpendicular direction ($\Omega_z$). 
Note that, by symmetry only $U_r \ne 0$ when $\theta=0$ or $\pi/2$, \ie, the passive particle is simply attracted towards or repelled from the active one; however, all three velocities components are nonzero when $\theta=\pi/4$.
Comparing our results with the far-field approximation shows that the two agree reasonably well even at small separations, particularly when $\theta=\pi/4$, consistent with the simulations of \citet{ishikawa2006hydrodynamic}.
At large separations, we also observe the expected scaling laws estimated above.

\subsection{Pair interactions}
\label{subsec:pair}

\begin{figure*}[t]
  \centering
  \includegraphics[height=7.5cm]{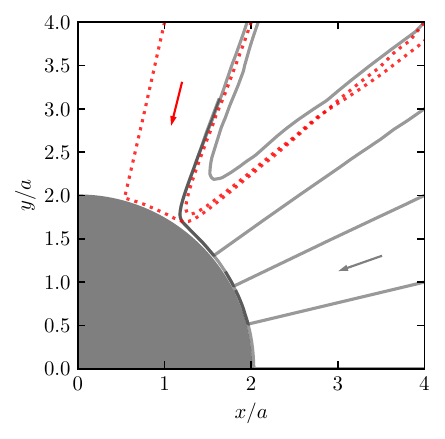} 
  \includegraphics[height=7.5cm]{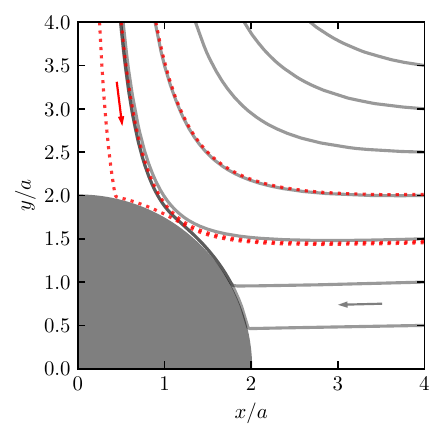} 
  \begin{picture}(0,0)
    \put(-15.5,7){(a)} \put(-7.8,7){(b)}
  \end{picture} 
  \caption{Relative trajectories between a reference shaker (pulling or pushing) and 
                a passive particle (a) or another \textit{aligned} shaker of the same kind (b).
                Gray solid lines are for pullers, and red dotted lines pushers.
                In both cases, the trajectory of the second particle is plotted in the frame co-rotating with the reference particle
                such that the reference particle is oriented horizontally.}
  \label{fig:pair-phase}
\end{figure*}
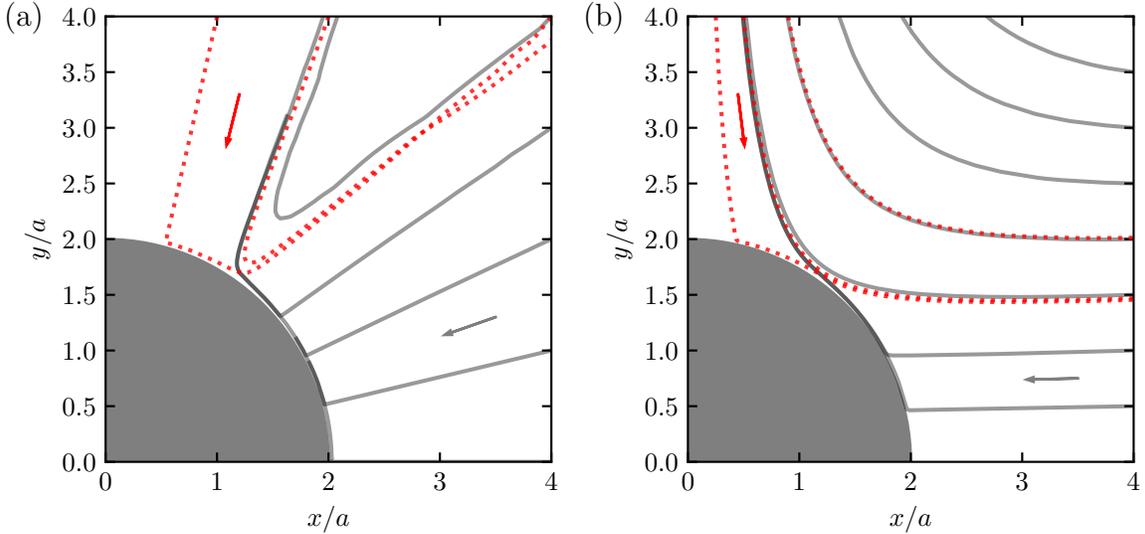

We can integrate the particle velocities in time to obtain a physical picture of their pair interactions.
Figure \ref{fig:pair-phase}(a) shows the trajectories of a passive particle next to a shaker in a frame \emph{co-rotating} with the reference shaker.
For a puller-type shaker, the region $|x|/a \lesssim 1$ cannot be reached from $|x|/a \gtrsim 1$ due to excluded-volume interactions (see Section \ref{sec:setup} for details of the repulsive force we impose), implying \emph{hysteretic} trajectories when reversing the boundary conditions on the shaker.
This is confirmed by the trajectories of the passive particle next to a pusher-type shaker (see red dotted lines), which intersect the gray solid lines in the figure.
Furthermore, these trajectories suggest that there is no stable configurations for a bound pair: depending on the initial position, the pair may temporarily approach each other but will always separate after the encounter, since all trajectories are open.
We note that there are two saddle points located at the poles ($x=\pm 2a$, $y=0$) of a pulling shaker, while there are infinitely many saddle points ($x=0$) along the equatorial rings of a pushing shaker.
The duration of the close encounters around those saddle points depends on the amount of angular noise in the dynamics.
In concentrated suspensions, we expect the lifetime of such transient pairs to be short due to the effect of many-body interactions.

Figure \ref{fig:pair-phase}(b) shows the relative trajectories between two identical, initially aligned shakers.
Similar to a shaker-passive pair, the trajectories are hysteretic (\ie, some regions are inaccessible) and all open (\ie, no bound pairs).
Under purely hydrodynamic interactions, an initially aligned pair will remain aligned due to the symmetries of the flow induced by a force dipole, because the rotation of the first shaker due to the second one is exactly the same as the rotation of the second shaker due to the first one, which we observe numerically.
However, we cannot expect all shakers to be aligned in a suspension because that implies \emph{nematic order}, which is absolutely unstable at long wavelengths in apolar suspensions \citep{Simha_Ramaswamy2002}.
Therefore, Figure \ref{fig:pair-phase}(b) provides some indication of how a pair of shakers interact, but does not represent a complete phase portrait.

\section{Results}
\label{sec:results}

\subsection{Simulation set-up}
\label{sec:setup}

\begin{table}[b]
  \centering
  \setlength{\tabcolsep}{14pt} \renewcommand{\arraystretch}{1.25} 
  \begin{tabular}{cccc}
    \hline
    $B_1$    &     $B_2$ $[a/\tau_a]$    &    $\phi$    &    $f_p$    \\
    \hline
    0     &   $-0.2$ (pushers), 0 (passive), or 0.2 (pullers)  &  0.5\% to 55\%    &    0 to 97\%   \\  
    \hline
  \end{tabular}
  \caption{Summary of the governing parameters. }
  \label{tab:param}
\end{table}

We simulate suspensions of shakers at various volume fractions ($\phi$) ranging from dilute ($\phi =0.5\%$) to dense ($\phi =55\%$).
At certain $\phi$, we have incorporated passive particles into the suspensions at varying ratios ($f_p$): $f_p=0$ represents a fully active suspension, while $f_p>0$ represents binary active suspensions.
Since we consider shaker-type squirmers, $B_1$ is always zero, but $B_2$ can be either positive (pullers) or negative (pushers).
Both $B_1$ and $B_2$ have the dimension of velocity, and at leading order the magnitude of $B_2$ determines the rate of the hydrodynamic interactions due to particle activity, as mentioned earlier. 
Throughout this work we fix $|B_2|$ and use it to define a reference time, $\tau_a \equiv a/|B_2|$, where $a$ is the particle radius.
This allows us to define the units in combination of $a$ and $\tau_a$ for all physical quantities involving length or time; \eg $a/\tau_a$ for translational velocity, $1/\tau_a$ for rotational velociy, $a^2/\tau_a$ for translational diffusivity, etc.
The results in this section will be reported in these units unless otherwise stated.
Finally, the values of the governing parameters are summarized in Table \ref{tab:param}.

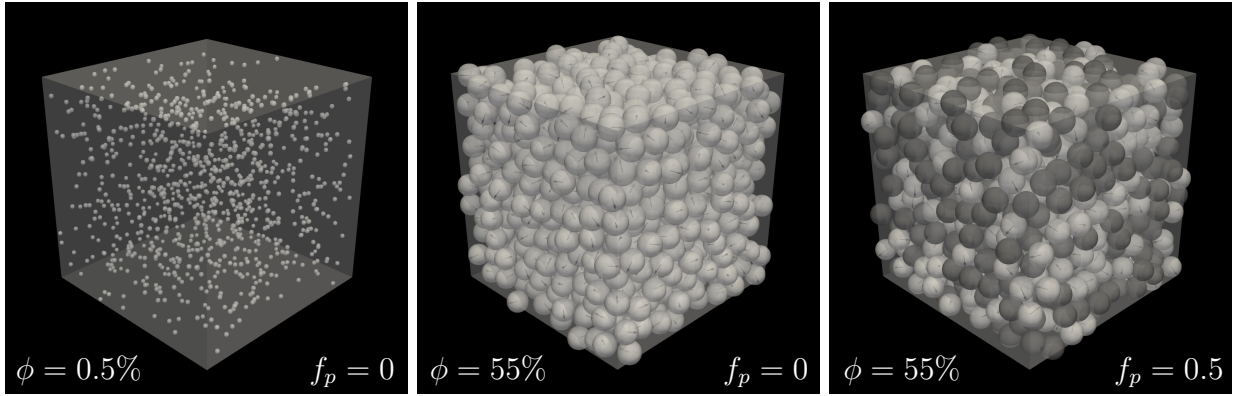
\begin{figure*}[t]
  \centering
  \includegraphics[height=5.2cm]{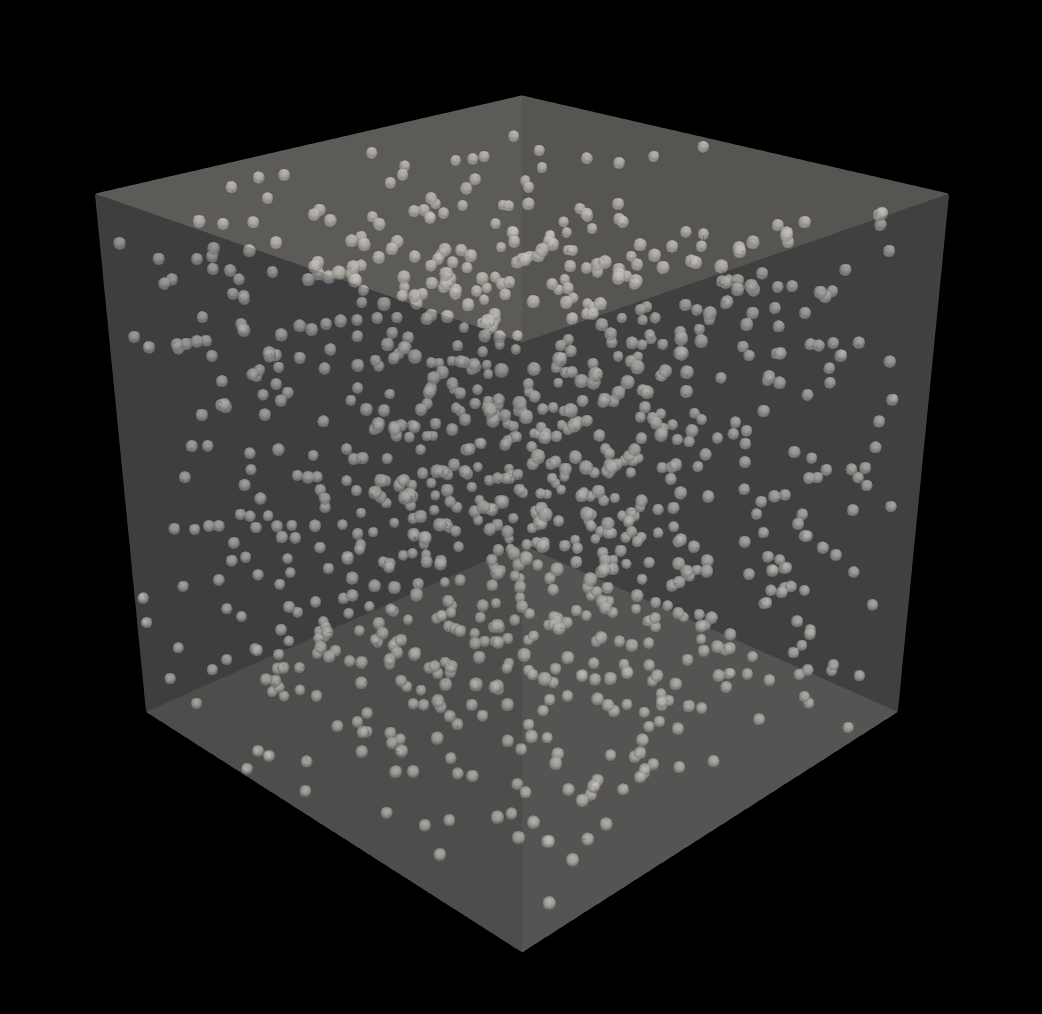} 
  \includegraphics[height=5.2cm]{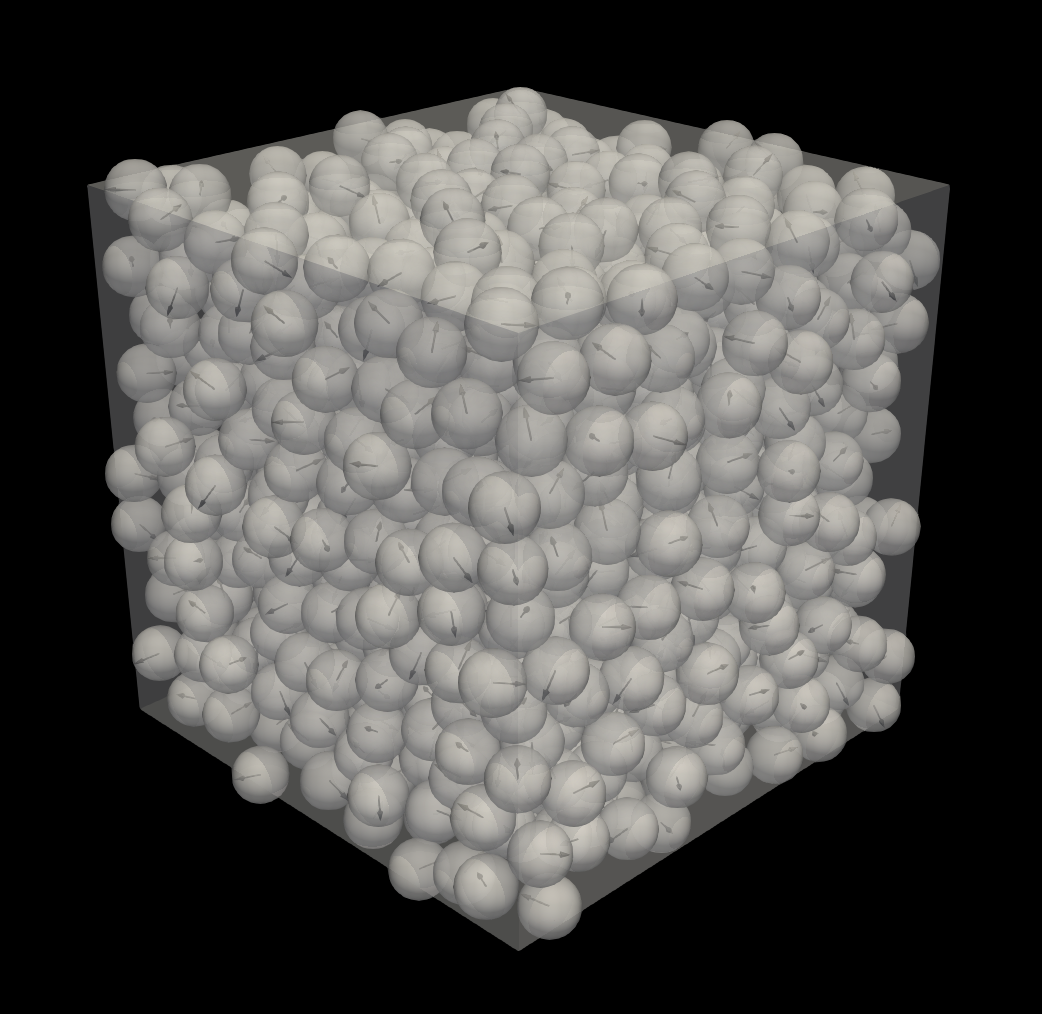} 
  \includegraphics[height=5.2cm]{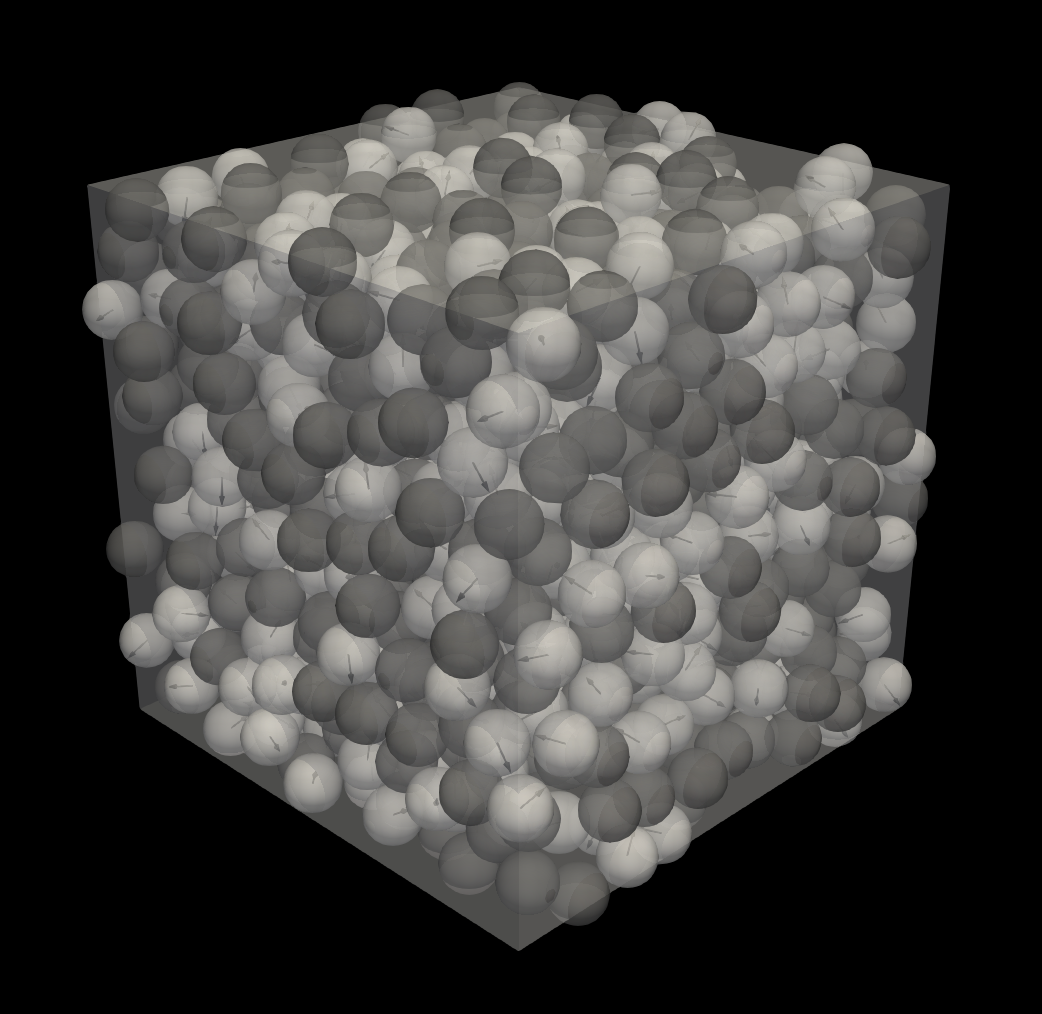} 
  \begin{picture}(0,0)
    \put(-16.3,0.2){\textcolor{white}{$\phi=0.5\%$}} 
    \put(-12.4,0.2){\textcolor{white}{$f_p=0$}} 
    \put(-10.8,0.2){\textcolor{white}{$\phi=55\%$}} 
    \put(-6.9,0.2){\textcolor{white}{$f_p=0$}} 
    \put(-5.3,0.2){\textcolor{white}{$\phi=55\%$}} 
    \put(-1.7,0.2){\textcolor{white}{$f_p=0.5$}} 
  \end{picture} 
  \caption{Suspensions of 1024 particles at different $\phi$ and $f_p$ (darker particles are passive).}
  \label{fig:snaps}
\end{figure*}

Numerically, we simulate $N=1024$ particles in cubic domains with periodic boundary conditions; see Figure~\ref{fig:snaps} and supplemental videos for illustration.
We have checked $N=2048$ and 4096, but observe no qualitative difference in the statistical results.
The initial condition, in terms of the particle position and orientation, is random, and three realizations of each case (in $B_2$ and $\phi$) are run for 100$\tau_a$ to reach the average steady state of the short-time dynamics (see Section \ref{sec:short}).
For suspension simulations, it is customary to impose a short-range repulsive force to prevent the particles from overlapping.
Here, we use the electrostatic repulsion typical for colloids, ${F}_\textrm{ext} = F_0 \exp(-\kappa h)$, where ${F}_0$ is the maximal repulsion, $\kappa$ the inverse Debye length, and $h$ the surface gap between two particles \citep{mewis_wagner_book}.
This introduces an additional time scale, $\tau_{e} \equiv 6\pi\eta a^2/F_0$, which should be smaller than $\tau_a$ but bigger than the numerical timestep ($\eta$ is the fluid dynamic viscosity).
As in our previous work \citep{Ge2022}, we choose $\tau_{e}/\tau_a=0.02$ and $\kappa^{-1}/a=0.01$ to model a strong and short-range repulsion.
The numerical timestep ($\Delta t$) is 100 times smaller than $\tau_{e}$, \ie, $\Delta t/ \tau_a = 2\times 10^{-4}$.

\subsection{Short-time dynamics: Instantaneous speeds}
\label{sec:short}

\begin{figure*}[t]
  \centering
  \includegraphics[width=\textwidth]{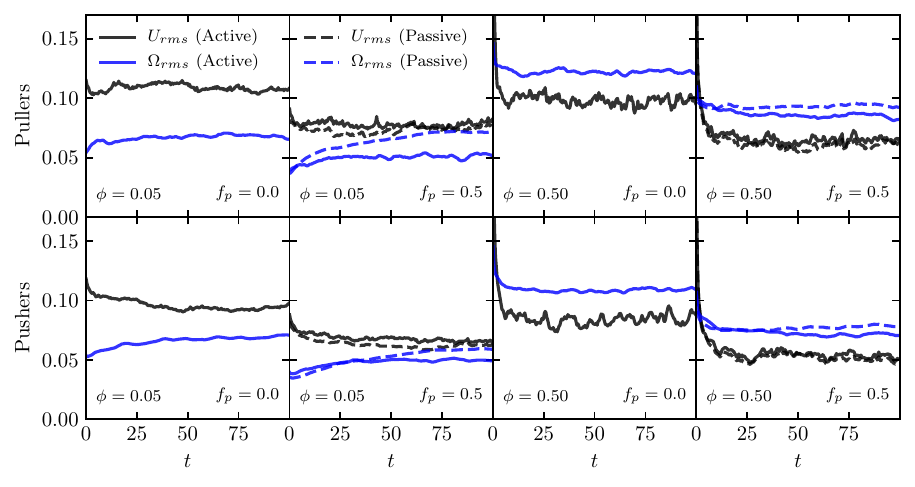}
  \caption{Instantaneous speeds of suspensions of pullers or pushers at $\phi=0.05$ or 0.5.
                In each case, a binary suspension with half passive particles ($f_p=0.5$) is also shown.}
  \label{fig:rms-time}
\end{figure*}

We first examine the short-time suspension dynamics, quantified by the root mean square (rms) particle speed.
Figure \ref{fig:rms-time} shows a few examples of the translational ($U_{rms}$) and rotational ($\Omega_{rms}$) rms speeds in time for dilute and dense suspensions with or without passive particles.
When $f_p=0$, both $U_{rms}$ and $\Omega_{rms}$ reach steady state in a relatively short time; however, the transients can be longer if the ratio of passive particles is higher or the volume fraction of the suspension is lower.
The longer transients are better seen in the evolution of $\Omega_{rms}$, which can also differ for shakers and passive particles.
The difference stems from the \emph{nonreciprocity} of hydrodynamic interactions.
In our active suspensions of smooth spheres, the only mechanism to transfer angular momentum is via hydrodynamic interactions. 
From Eq.~\eqref{eq:U_p}, we can infer that the rotational motion of a passive particle due to a nearby shaker is larger than the converse, because the flow induced by a passive particle is of higher order.
Therefore, unless the interactions cancel out, passive particles will have a larger $\Omega_{rms}$.
On the other hand, the translational momentum can be transferred through both hydrodynamic and excluded-volume interactions, where the latter tends to homogenize the momentum distribution.
Consequently, the nearly equal $U_{rms}$ among shakers and passive particles suggests the importance of excluded-volume interactions for the short-time dynamics.
This is observed in most of the cases that we simulated, except when $\phi(1-f_p)$ is small, in which the dynamics are slow because the concentration of active particles is low.

\begin{figure*}[t]
  \centering
  \includegraphics[width=\textwidth]{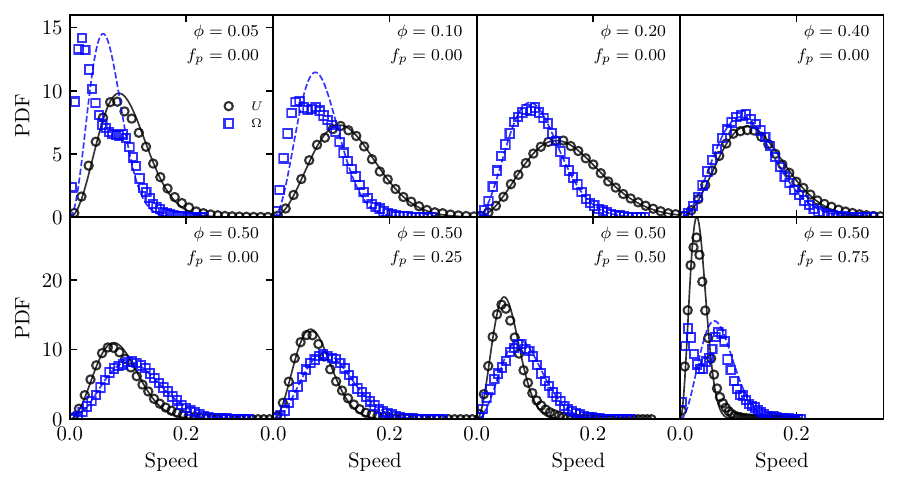}
  \caption{Distribution of particle speeds at different $\phi$ and $f_p$ for suspensions of pulling shakers and passive particles.
                Lines are fits to the Maxwell-Boltzmann distribution.}
  \label{fig:speed-distr}
\end{figure*}

The short-time dynamics can be analyzed further by examining the distribution of particle speeds.
Figure \ref{fig:speed-distr} shows the probability density functions (PDFs) of the translational ($U$) and rotational ($\Omega$) speeds sampled over all particles at steady state.
Here, we do not distinguish between active and passive particles because their distributions have nearly the same shape when sampled over a long time.
Several interesting observations can be made from these PDFs.
First, as found in previous studies \citep{WL2000, Patteson2016}, the translational speeds are always well-fitted by the Maxwell-Boltzmann (MB) distribution, which describes the behavior of fluids at equilibrium, even though our suspensions are highly dissipative and far from equilibrium.
Second, although the angular speeds tend to be MB when $\phi(1-f_p)$ is high, it can also be \emph{bimodal} in the opposite limit.
The bimodal distribution suggests that there are two characteristic speeds in the suspension, yet both must result from hydrodynamic interactions, which only have one characteristic time scale ($\tau_a$).
Third, the relative magnitudes of the characteristic translational and rotational speeds seem to vary non-monotonically with $\phi$, though both of them tend to reduce with $f_p$ (see Appendix \ref{sec:fp}).

We can explain these observations as follows.
Consider any test particle in a suspension of shakers.
Although no particle can move on its own, the flow induced by each active particle will generate a mean flow that advects all particles suspended in it.
If the suspension is \emph{homogeneous}, the average number of active neighbors per unit length at distance $r$ will be $n \sim 4\pi r^2\phi_a$, where $\phi_a \equiv \phi (1-f_p)$.
Furthermore, if the particle orientations are \emph{isotropic}, the typical magnitude of the flow due to all particles at the same distance will scale as $\sqrt{n} /r^2 \sim \sqrt{\phi_a}/r$. 
From Eqs.~(\ref{eq:U_p},\ref{eq:u_sol}), we thus expect the typical translational and rotational speeds due to particles at distance $r$ to scale as $\sqrt{\phi_a}/r$ and $\sqrt{\phi_a}/r^2$, respectively.
As these are \emph{statistical} estimates, the scalings are only valid at large distances. 
However, it is clear that the variances of both speeds are finite because the particle size is finite.
Consequently, the distributions of the net velocities upon integrating the velocity contributions over all distances tend to be Gaussian per the central limit theorem.
Furthermore, since there is no reason for any of the velocity components to differ from one another (otherwise the suspension would not be isotropic), their speeds necessarily display the MB distribution. 
These conditions are verified in Appendix \ref{sec:nematic}.
On the other hand, if $\phi_a$ is small, the convergence towards Gaussian will be slow and both $U$ and $\Omega$ can deviate from MB.
We expect more deviation in $\Omega$ because ${\bm\Omega}_p$ decays faster in $r$ than ${\bm U}_p$.
Hence, the bimodal distribution observed for $\Omega$ is a statistical effect;
the additional peak at low speeds in the PDFs corresponds to particles without sufficient active neighbors within a large distance.

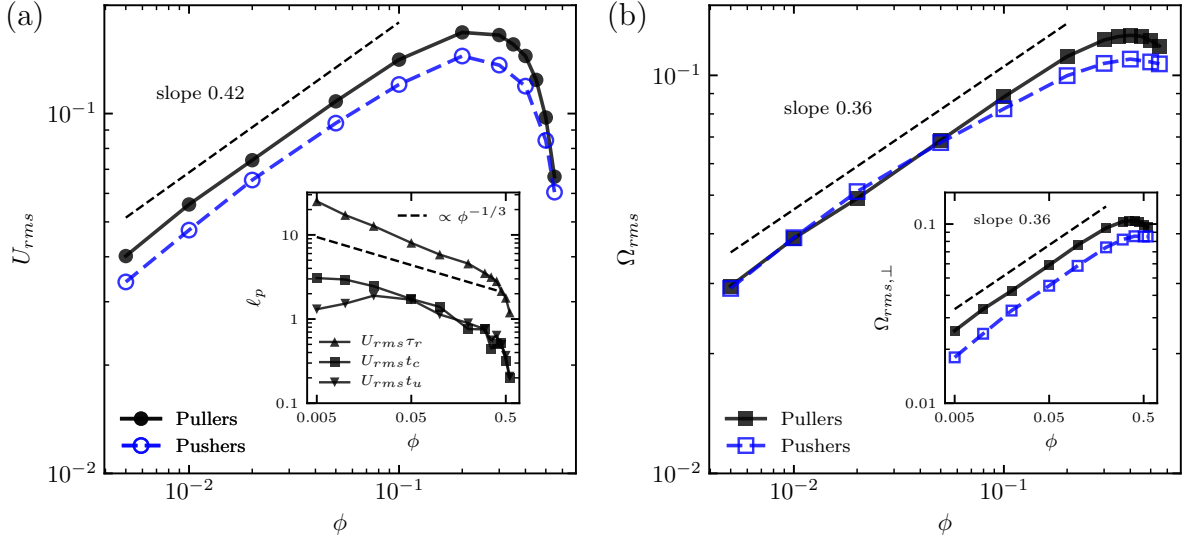
\begin{figure*}[t]
  \centering
  \includegraphics[height=7.5cm]{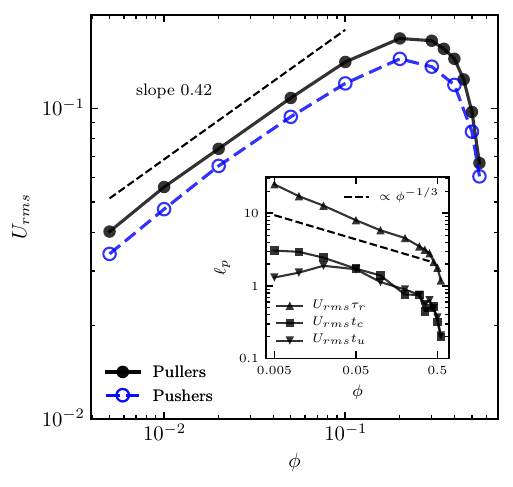} 
  \includegraphics[height=7.5cm]{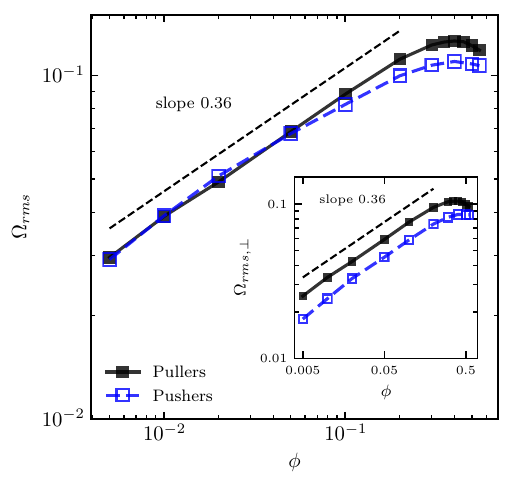} 
  \begin{picture}(0,0)
    \put(-16.,7){(a)} \put(-8.,7){(b)}
  \end{picture} 
  \caption{Typical instantaneous speeds of the translational (a) and rotational (b) motion 
                in suspensions of pullers or pushers at different volume fractions.
                The inset in (a) shows the persistence length in suspensions of pullers, 
                while the inset in (b) shows the perpendicular component of $\Omega_{rms}$; 
                see Section \ref{sec:long} for details.
                In both (a) and (b), $f_p=0$.}
  \label{fig:rms-phi}
\end{figure*}

Figure \ref{fig:rms-phi} shows the rms speeds of suspensions of pullers or pushers at different volume fractions.
The results are generally similar for pullers and pushers, though pullers tend to move slightly faster than pushers. 
When $\phi < 0.2$, both $U_{rms}$ and $\Omega_{rms}$ increase with $\phi$, but the scaling exponents are less than the expected $1/2$  based on the statistical analysis valid at large distances. 
Therefore, the velocity contributions from nearby particles must have a weaker dependence on $\phi$, which we verify in Section \ref{sec:correlations}.
Another observation is that these speeds do not increase monotonically with $\phi$.
The non-monotonic behavior of $U_{rms}$ is expected from the competition of hydrodynamic and excluded-volume interactions.
In dilute suspensions, the mean kinetic energy of the translational motion increases with $\phi$ because \hi increase with $\phi$.
However, as $\phi$ approaches the jamming volume fraction, the dynamics necessarily slow down due to the reduction of free space between particles.
If so, the persistence length ($\ell_p$) of the ballistic motion may reduce with the volume fraction as $\ell_p \sim \phi^{-1/3}$.
The inset in Figure \ref{fig:rms-phi}(a) shows three estimations of $\ell_p$, all of which are roughly consistent with the $\phi^{-1/3}$ scaling until $\phi \approx 0.3$.
At still higher $\phi$, the mean distance between neighboring particles becomes comparable to two particle radii and $\ell_p$ reduces sharply.
It is in this regime that $U_{rms}$ decreases with $\phi$.
We shall defer the definitions of $\tau_r$, $t_{c}$ and $t_{u}$ to the next section (\ref{sec:long}) because they are related to the long-time dynamics; for now, it suffices to note that the different measures give similar results.

Finally, the slight non-monotonic behavior of $\Omega_{rms}$ is intriguing. 
From the reasoning above, we would expect $\Omega_{rms}$ to increase monotonically with $\phi$, as excluded-volume interactions do not affect the rotation of spheres and the pairwise \hi are stronger when particles are closer.
Therefore, the slight decrease of $\Omega_{rms}$ when $\phi > 0.4$ may suggest a partial cancellation of the hydrodynamic effect on particle rotation.
Since \hi are determined by the relative positions and orientations of all particles, this further implies that there may be certain \emph{correlations} in the suspension microstructure, particularly in the dense regime.
We will analyze such correlations in detail in Section \ref{sec:correlations}.

\subsection{Long-time dynamics: Hydrodynamic diffusion}
\label{sec:long}

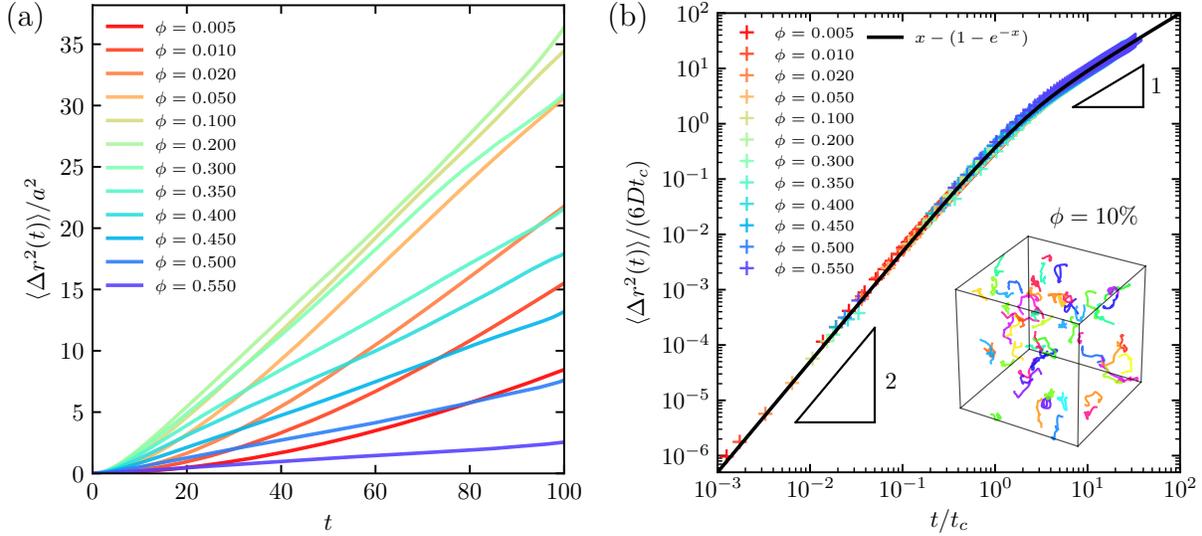
\begin{figure*}
  \centering
  \includegraphics[height=7.5cm]{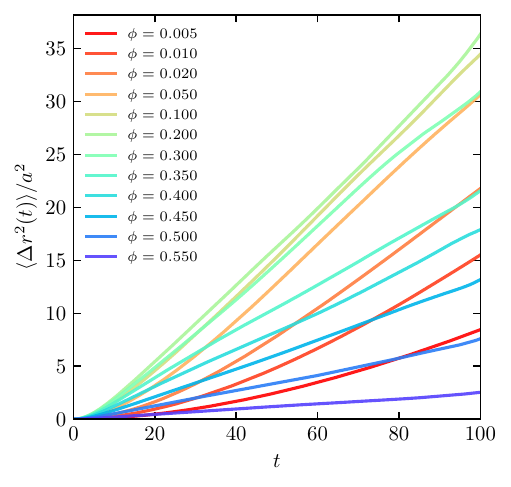}
  \includegraphics[height=7.5cm]{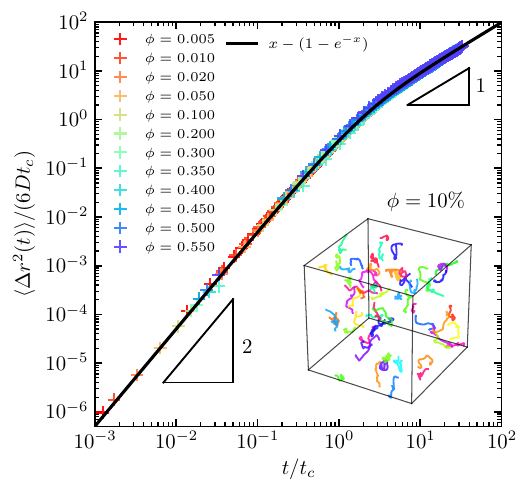}
  \begin{picture}(0,0)
    \put(-16.2,7){(a)} \put(-8.2,7){(b)}
  \end{picture} 
  \caption{(a) MSD in time at different volume fractions for suspensions of pulling shakers at $f_p=0$.
               (b) The same data rescaled according to Eq.~\eqref{eq:wl2000}, where the inset shows sample particle trajectories at $\phi=10\%$.}
  \label{fig:msd}
\end{figure*}

To characterize the long-time dynamics of our suspensions, we first calculate the mean square displacement (MSD) of the particles, defined as $\langle \Delta r^2(t) \rangle \equiv \big\langle [{\bm r}(\tau + t) - {\bm r}(\tau)]^2 \big\rangle_\tau$, where ${\bm r}(t)$ is the position of a particle at time $t$, and $\langle \cdot \rangle_\tau$ is an average over all particles and reference times $\tau$.
Figure \ref{fig:msd}(a) shows the temporal evolution of the MSDs for suspensions of pulling shakers at different volume fractions.
The nearly linear growth of the MSDs signifies a diffusive process, which has indeed been observed in various active systems \citep{WL2000, Leptos2009, Patteson2016, Peng2016}.
For bacterial suspensions, \citet{WL2000} (WL) developed a simple stochastic model to analyze the diffusion of immersed passive beads.
Assuming a ``collisional force" that is exponentially correlated in time between the beads (\ie, the dynamics of the beads are not random but correlated), it can be shown that the MSD in three dimensions is governed by
\begin{equation} \label{eq:wl2000}
  \langle \Delta r^2(t) \rangle = 6D \big[t - t_{c} \big(1- e^{-t/t_{c}} \big) \big],
\end{equation}
where $D$ is the effective diffusion coefficient, and $t_{c}$ is the lifetime of the coherent structures in the flow.
Fitting our data according to Eq.~\eqref{eq:wl2000} collapses all the results, as illustrated in Figure \ref{fig:msd}(b), suggesting that the interactions between the shakers may be effectively described by the WL model \footnote{In the original WL model, the MSD is expressed as $\langle \Delta r^2(t) \rangle = 4D t[1- \exp(-t/t_{c})]$, which is in two dimensions and neglects a short-time contribution. The full expression, Eq.~\eqref{eq:wl2000}, fits our data better.}.
Furthermore, the fitting reveals that the dynamics are ballistic in the short times and only become diffusive after $t>t_{c}$.

\begin{figure*}
  \centering
  \includegraphics[height=7.5cm]{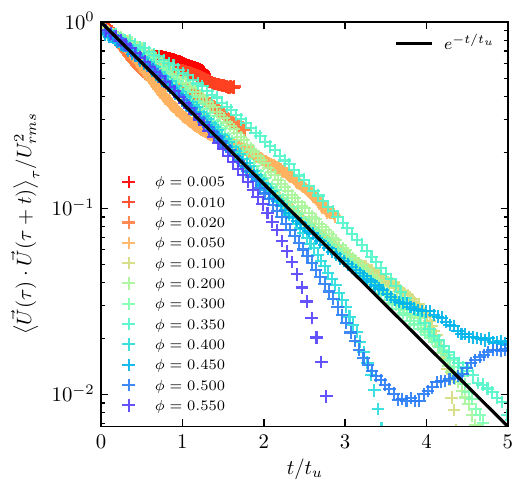}
  \includegraphics[height=7.5cm]{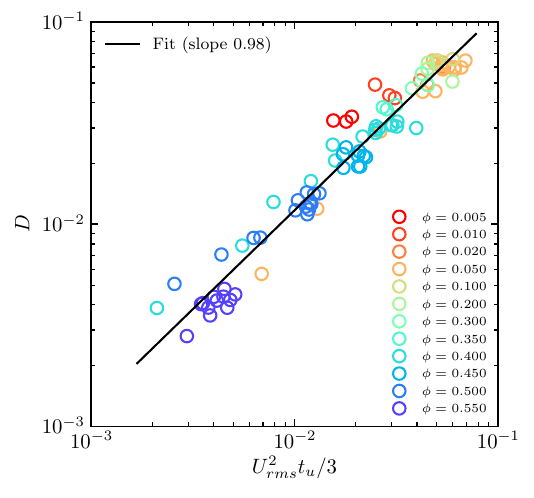}
  \begin{picture}(0,0)
    \put(-16,7){(a)} \put(-8.2,7){(b)}
  \end{picture} 
  \caption{(a) Velocity autocorrelation in time at different volume fractions for suspensions of pulling shakers at $f_p=0$.
               (b) The scaling between $D$ and $U_{rms}^2 t_{u}$ for all simulated cases.}
  \label{fig:vdv}
\end{figure*}
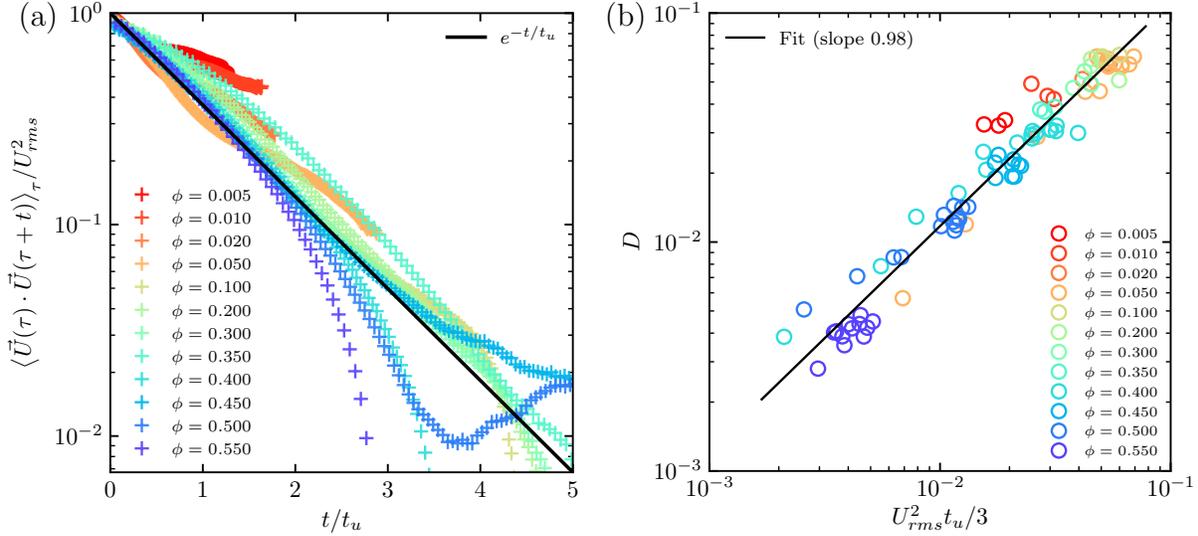

To further verify the WL model against our data, we fitted the velocity autocorrelation, $\langle {\bm U}(\tau) \cdot {\bm U}(\tau +t) \rangle_\tau$, according to 
\begin{equation} \label{eq:wl2000vdv}
  \langle {\bm U}(\tau) \cdot {\bm U}(\tau +t) \rangle_\tau = U_{rms}^2 e^{-t/t_{u}},
\end{equation}
where ${\bm U}(\tau)$ is the velocity of a particle at time $\tau$, $\langle \cdot \rangle_\tau$ is an average over all particles and reference times $\tau$, and $t_{u}$ is the characteristic time for the velocity decorrelation.
The results of the fits are shown in Figure \ref{fig:vdv}(a).
If the dynamics of our suspensions are correctly described by the WL model, the two time scales, $t_{c}$ and $t_{u}$, should be equal \citep{WL2000}.
Indeed, we observe $t_{c} \approx t_{u}$ at nearly all volume fractions except when $\phi$ is very small; c.f.~Figure \ref{fig:rms-phi}(a) inset.
When $\phi \leqslant 0.01$, the dynamics are too slow for the velocity statistics to converge at the end of the simulations, thus $t_{u}$ is less reliable than  $t_{c}$, as can also be seen in the quality of the fit in Figure \ref{fig:vdv}(a).
Nevertheless, the scaling between the instantaneous and long-time dynamics are consistent, and we verify that $D \approx U_{rms}^2 t_{u}/3$ as expected from the Taylor-Green-Kubo relation \citep{graham2018book} and the exponential decay of the velocity autocorrelation, c.f.~Figure \ref{fig:vdv}(b).
We have checked that the scalings for pushers are the same.
These comparisons thus cross-validate the WL model and our results.
In the following, we will use Eq.~\eqref{eq:wl2000} to extract the diffusion coefficient ($D$) and crossover time ($t_c$) of the translational dynamics.

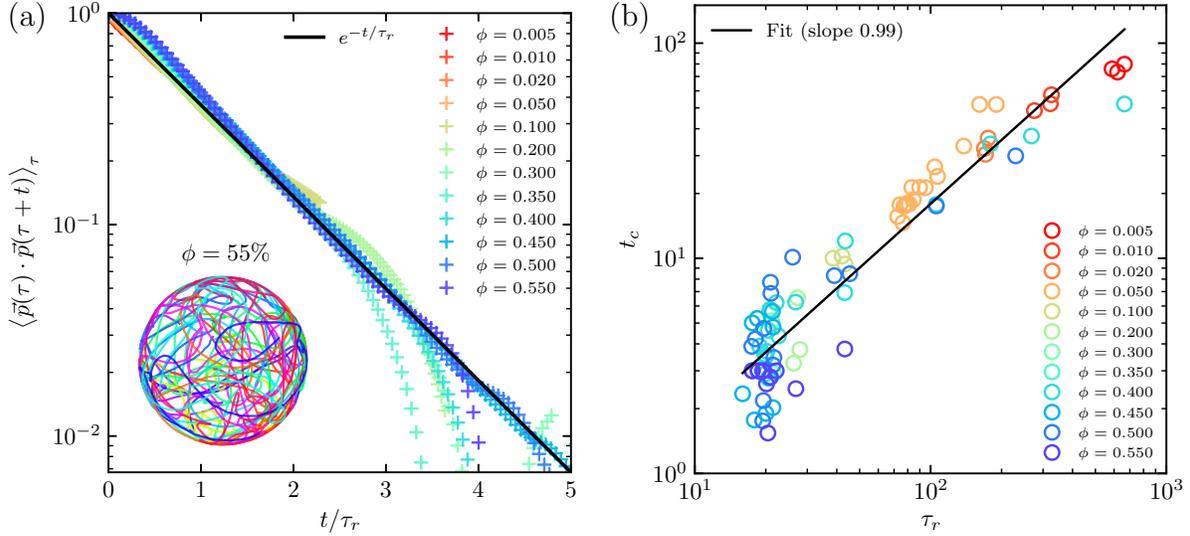
\begin{figure*}
  \centering
  \includegraphics[height=7.5cm]{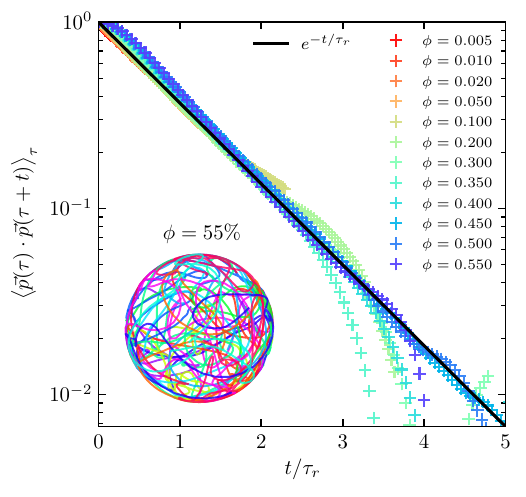}
  \includegraphics[height=7.5cm]{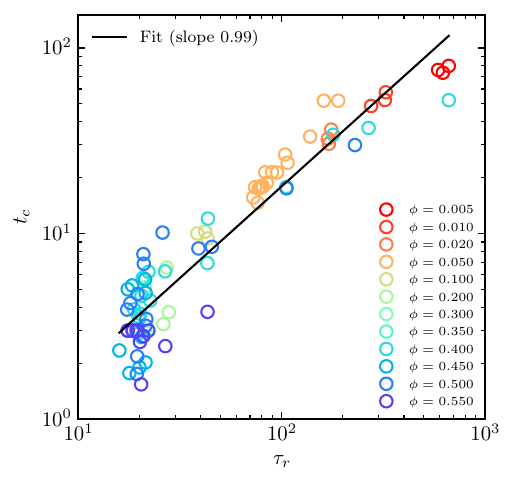}
  \begin{picture}(0,0)
    \put(-16,7){(a)} \put(-8,7){(b)}
  \end{picture} 
  \caption{(a) Orientation autocorrelation in time at different volume fractions for suspensions of pulling shakers at $f_p=0$;
                inset shows sample orientations traced by the particles on a unit sphere at $\phi=55\%$. 
                (b) The scaling between $t_c$ and $\tau_r$ for all cases.}
  \label{fig:ndn}
\end{figure*}

The long-time rotational dynamics can be quantified by the relaxation of the particle orientations.
Specifically, we calculate the autocorrelation function, $\langle {\bm p}(\tau) \cdot {\bm p}(\tau +t) \rangle_\tau$, and fit it according to
\begin{equation} \label{eq:ndn}
  \langle {\bm p}(\tau) \cdot {\bm p}(\tau +t) \rangle_\tau = e^{-t/\tau_r},
\end{equation}
where ${\bm p}(\tau)$ is the orientation of a particle at time $\tau$, $\langle \cdot \rangle_\tau$ is an average over all particles and reference times $\tau$, and $\tau_r$ is the characteristic time of the orientation decorrelation.
Figure \ref{fig:ndn}(a) shows that $\langle {\bm p}(\tau) \cdot {\bm p}(\tau +t) \rangle_\tau$ indeed decays exponentially in time at all volume fractions.
For illustration, we also plot the ``trajectories" traced by the particle axes, which can be in any direction and changes smoothly on a unit sphere.
Physically, we expect $\tau_r$ to be similar to $t_c$, as both are related to the persistence of the ballistic motion.
Figure \ref{fig:ndn}(b) shows that, despite the scatter, $t_c \propto \tau_r$ and $t_c < \tau_r$.
Therefore, the translational dynamics cease to be ballistic before the particle orientations are fully randomized by the rotational diffusion, because the former is also affected by excluded-volume interactions.

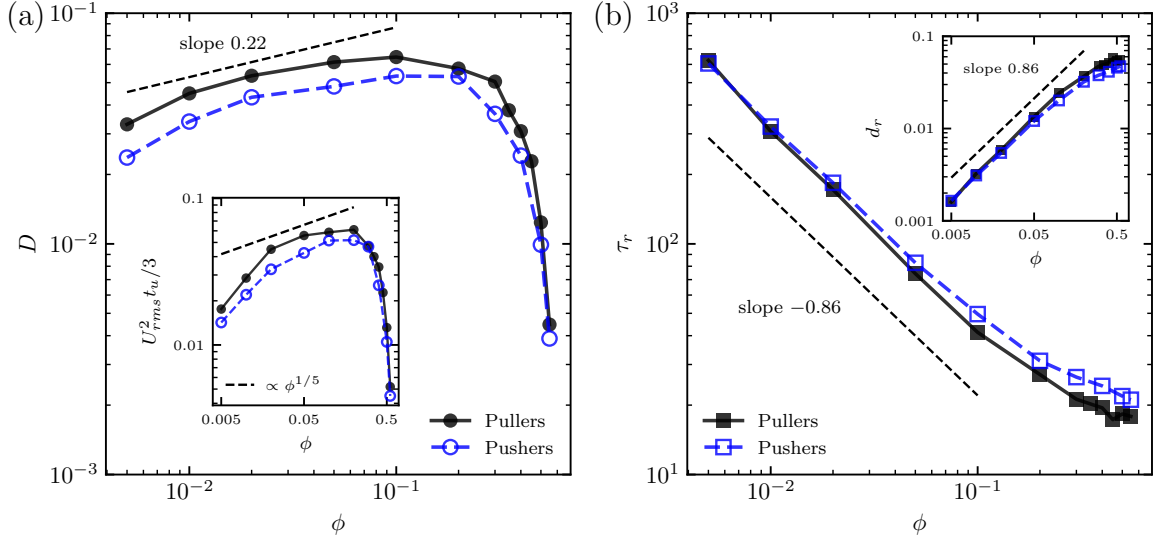
\begin{figure*}
  \centering
  \includegraphics[height=7.5cm]{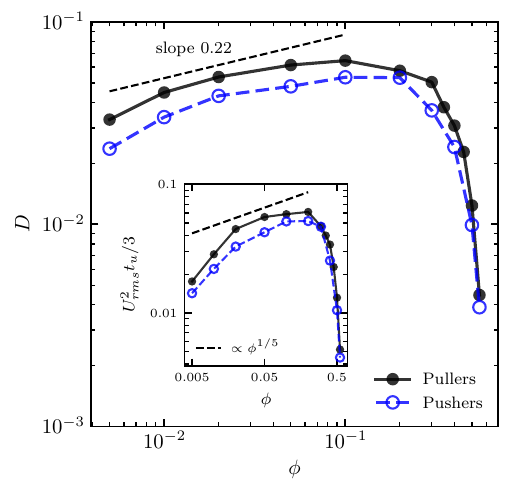}
  \includegraphics[height=7.5cm]{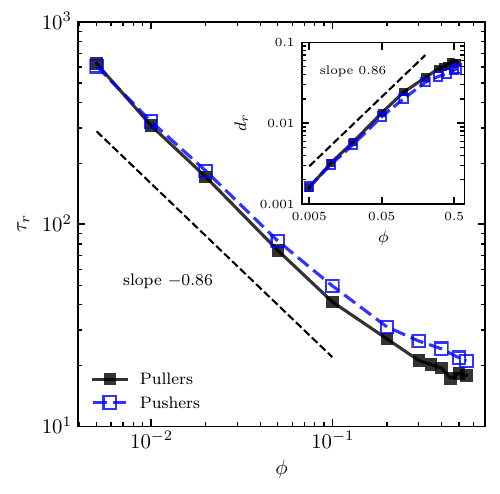}
  \begin{picture}(0,0)
    \put(-15.6,7){(a)} \put(-7.8,7){(b)}
  \end{picture} 
  \caption{Hydrodynamic diffusion.
               (a) Translational diffusion coefficient for suspensions of pullers or pushers at different volume fractions;
               inset shows the diffusivity estimated from short-time dynamics.
               (b) Rotational relaxation time for suspensions of pulling shakers;
               inset shows the rotational diffusivity $d_r$ vs $\phi$. 
               In both (a) and (b), $f_p=0$.}
  \label{fig:diffusion}
\end{figure*}

The fitting procedures presented above allow us to examine the long-time translational and rotational dynamics, characterized by $D$ and $\tau_r$, respectively, at different $\phi$ and $f_p$.
Figure \ref{fig:diffusion} shows that, as for the short-time dynamics, the results of pullers and pushers are similar (with pullers being slightly more diffusive) and $D$ is clearly non-monotonic in $\phi$.
The dependence of $D$ on $\phi$ resembles that of $U_{rms}$ and can be explained by the same underlying mechanism (hydrodynamic vs excluded-volume interactions); however, we note that the peak $D$ occurs at a lower $\phi$ than the peak $U_{rms}$.
This is due to the hydrodynamic coupling of the translational and rotational motion.
Specifically, because $D \sim U_{rms}^2 t_u \sim U_{rms}^2 \tau_r$, its derivative $\frac{\partial D}{\partial \phi} \sim D(\frac{2}{U_{rms}}\frac{\partial U_{rms}}{\partial \phi} + \frac{1}{\tau_r}\frac{\partial \tau_r}{\partial \phi})$; and since $\tau_r$ decreases with $\phi$, the maximum $D$ is reached while $U_{rms}$ is still increasing.
For the rotational dynamics, the inverse of $\tau_r$ is sometimes used to define a rotational diffusion coefficient, $d_r \equiv \tau_r^{-1}$ \citep{Saintillan_Shelley2007}.
Figure \ref{fig:diffusion}(b) shows that $d_r$ increases sublinearly with $\phi$ when $\phi \lesssim 0.1$.
This is roughly consistent with what we expect from the short-time dynamics, as can be verified with $\ell_p \sim U_{rms} t_c \sim \phi^{-1/3}$ and $\tau_r \sim t_c$; c.f.~Figures \ref{fig:rms-phi}(a) and \ref{fig:ndn}(b).
We note that in the literature it is sometimes reported that $D \sim \phi^{-1}$ and $d_r \sim \phi$ \citep{ishikawa2007diffusion, Saintillan_Shelley2007}.
A key difference between our system and those mentioned lies in the propulsion mechanism.
The previous works considered self-propelling particles ($U_{0} \sim$ const.), thus the persistence length is determined by the mean free path, $\ell_{0} \sim \phi^{-1}$, leading to $D \sim U_{0} \ell_{0} \sim \phi^{-1}$ and $d_r \sim \tau_r^{-1} \sim U_{0}/\ell_{0} \sim \phi$.
In our case, there is no self-propulsion and the translational motion ($U_{rms}$) is fully coupled to the particle rotation ($\tau_r$); hence, the different scalings.

A comparison between Figures \ref{fig:rms-phi}(b) and \ref{fig:diffusion}(b) shows that $d_r$ does not reduce as in the case of $\Omega_{rms}$ at higher $\phi$.
One possible reason is that particles may rotate about their own axes, thus leaving their orientation unchanged. 
However, we have checked the perpendicular component of the angular speed and find it to exhibit the same $\phi$ dependence as the full speed; see the inset in Figure \ref{fig:rms-phi}(b).
The translational and rotational velocities of the particles are also uncorrelated to their orientations in our suspensions; \ie, $\langle {\bm U} \cdot {\bm p} \rangle \approx 0$ and $\langle {\bm \Omega} \cdot {\bm p} \rangle \approx 0$.
We are uncertain about the source of this minor discrepancy.

Finally, we have also examined the long-time dynamics in binary suspensions; \ie, $f_p>0$.
In general, both $D$ and $d_r$ tend to reduce with $f_p$ at a fixed $\phi$ as one would expect; however, the reduction of $D$ is nonlinear and, in some cases, it may even increase slightly with $f_p$ when $f_p \ll 1$.
A brief discussion of the $f_p$ dependence is provided in Appendix \ref{sec:fp}.

\subsection{Microstructure: Spatial correlations}
\label{sec:correlations}

From the short-time dynamics we inferred that certain spatial correlations, particularly over short distances, may be present in our suspensions.
Therefore, in the reminder of the paper, we examine the near-field microstructure in both particle position and orientation.

\begin{figure*}[t]
  \centering
  \includegraphics[width=\textwidth]{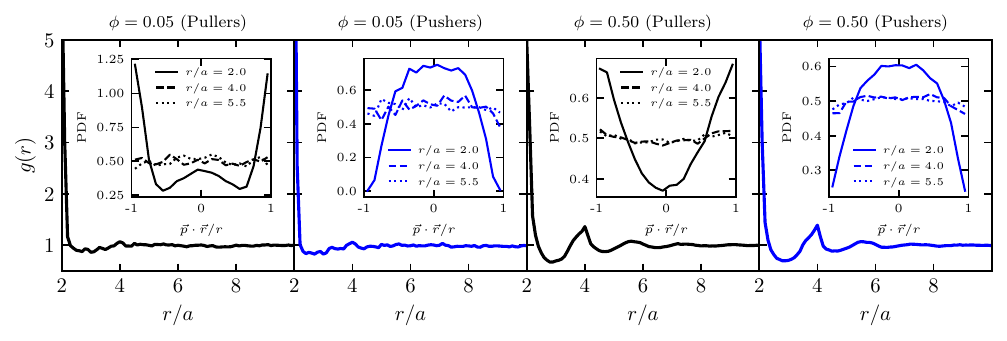}
  \caption{Suspension microstructure. 
                In each panel, the main plot shows the radial distribution function, $g(r)$, 
                in suspensions of pullers or pushers at $\phi=0.05$ and 0.5,
                whereas the inset corresponds to the angular distribution at $r/a=2$, 4, and 5.5.}
  \label{fig:rdf-adf}
\end{figure*}

We know that long-range nematic order is absolutely unstable in apolar active suspensions \citep{Simha_Ramaswamy2002}.
For shakers, this was shown to result in an isotropic global orientation distribution \citep{evans2011orientational}, which we confirm in our simulations; see Appendix \ref{sec:nematic}.
To examine the \emph{local} order, we calculate the pair correlation function, 
\begin{equation} \label{eq:g(r)}
g({\bm r}) = \frac{\big\langle \sum_{i \ne j}^N \delta({\bm r}-{\bm r}_{ij}) \big\rangle}{4\pi r^2\rho},
\end{equation}
where $\rho$ is the average particle number density, and $\langle \cdot \rangle$ is an average over all particles and time.
Figure \ref{fig:rdf-adf} shows a few examples of suspensions of pullers or pushers at different $\phi$.
Here, the radial distribution, $g(r)$, displays the typical oscillatory behavior (especially at higher $\phi$) with decaying peaks at $r/a \approx 2$, 4, 5.5, and so on, corresponding to the first few particle layers around any particle.
However, the angular distributions in the first particle layer are \emph{anisotropic}.
Specifically, pulling shakers tend to find their nearest neighbors along the particle axis (\ie, $|{\bm p}\cdot{\bm r}|/r \approx 1$), whereas pushing shakers tend to find them in the plane perpendicular to ${\bm p}$ (\ie, ${\bm p}\cdot{\bm r}/r \approx 0$); see the insets in Figure \ref{fig:rdf-adf}.
Furthermore, since the same correlation applies also to the neighbors, adjacent shakers are approximately \emph{aligned}: for pullers, the alignment is head-to-head; for pushers, it is side-by-side. 
However, we cannot visually identify any particle chains or sheets as the global microstructure remains disordered (see supplemental videos) and the angular distribution quickly becomes isotropic beyond the first particle layer.

\begin{figure*}
  \centering
  \includegraphics[height=7.4cm]{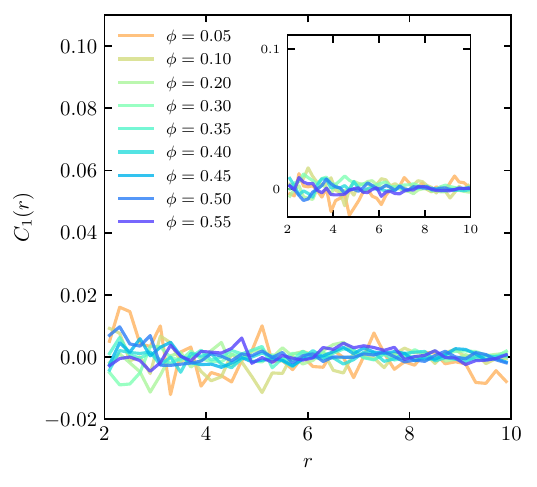}
  \includegraphics[height=7.4cm]{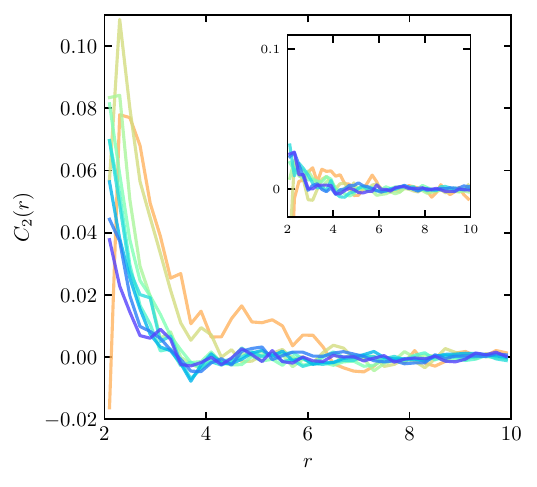}
  \begin{picture}(0,0)
  \end{picture} 
  \caption{Correlation functions for the particle orientations at $f_p=0$.
               (a) Polar order, $C_1(r)$. (b) Nematic order, $C_2(r)$.
               In both cases, the main plots correspond to suspensions of pullers, while the insets correspond to pushers (same legend).}
  \label{fig:C(r)}
\end{figure*}

To quantify the local alignment, we compute the following correlation functions for the particle orientation \citep{Saintillan_Shelley2007},
\begin{equation} \label{eq:C(r)}
  C_1(r) = \frac{\big\langle \sum_{i \ne j}^N ({\bm p}_i \cdot {\bm p}_j) \delta(r-r_{ij}) \big\rangle}{\big\langle \sum_{i \ne j}^N \delta(r-r_{ij})  \big\rangle} , \quad
  C_2(r) = \frac{\big\langle \sum_{i \ne j}^N \frac{1}{2}[3({\bm p}_i \cdot {\bm p}_j)^2-1] \delta(r-r_{ij}) \big\rangle}{\big\langle \sum_{i \ne j}^N \delta(r-r_{ij})  \big\rangle} ,
\end{equation}
where $\langle \cdot \rangle$ is an average over all particles and time.
Here, $C_1(r) \in [-1,1]$ measures the polar order, while $C_2(r) \in [-\frac{1}{2},1]$ measures the nematic order; both tend to be zero if there is no order. 
Figure \ref{fig:C(r)} shows that there is indeed a minor local nematic order, despite the absence of polar order at all distances.
For pullers, particles are more aligned at lower volume fractions, where the alignment can persist to $r/a \approx 4$.
For pushers, the alignment is generally weaker and appears to be insensitive to volume fraction.
Both of these observations are consistent with what we inferred from the angular distributions, c.f.~the insets of Figure \ref{fig:rdf-adf}.
Comparing to suspensions of self-locomoting rods \citep{Saintillan_Shelley2007, li2019data}, the degree of alignment in our suspensions of spherical shakers is much weaker (in \citet{Saintillan_Shelley2007}, $C_2(0) \approx 0.9$).
This suggests that particle shape or self-propulsion can have a large effect on the local order.

\begin{figure*}
  \centering
  \includegraphics[height=7.5cm]{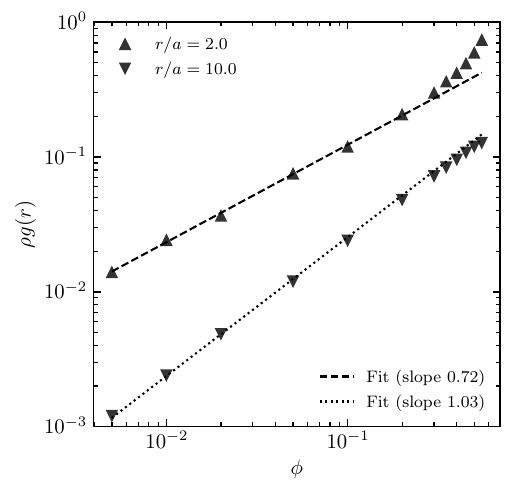}
  \includegraphics[height=7.5cm]{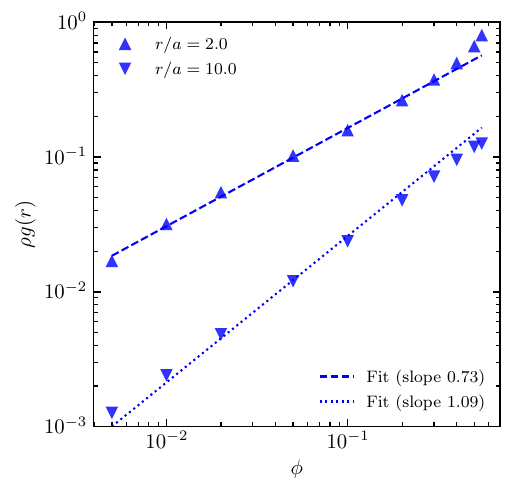}
  \begin{picture}(0,0)
  \end{picture} 
  \caption{Neighbor densities at small and large distances for suspensions of pullers (a) or pushers (b).
               In both cases, $f_p=0$ and lines are fits of the data for $\phi \leqslant 0.2$.}
  \label{fig:nb-phi}
\end{figure*}

Finally, we examine the scalings of the neighbor densities, proportional to $\rho g(r)$, at small and large distances.
Figure \ref{fig:nb-phi} shows that, for both pullers and pushers, $\rho g(r)$ scales linearly with $\phi$ at $r/a=10$, consistent with what we expect in an ideal homogeneous suspension; however, the scalings are \emph{sublinear} at $r/a=2$.
Therefore, the slower growth of $U_{rms}$ and $\Omega_{rms}$ with respect to $\phi$ (c.f.~Figure \ref{fig:rms-phi}) can be attributed to local correlations in the microstructure.
Such local correlations may also explain the slight reduction of $\Omega_{rms}$ or the weaker growth of $d_r$ when $\phi >0.4$, because particles tend to form layers at higher volume fractions and the effect of \hi from aligned neighboring particles on opposite sides cancel out.
However, since we can only quantify the ordering and alignment statistically, any cancellation is also partial and statistical.

\section{Conclusion}
\label{sec:conclusion}

In summary, we have presented a numerical study on the hydrodynamic diffusion of apolar active suspensions at volume fractions ($\phi$) ranging from 0.5\% to 55\%. 
We model the active particles as squirmers, with zero self-propulsion but non-zero self-straining, and simulate their collective dynamics using a recently developed active Fast Stokesian Dynamics method.
Our results show that both the instantaneous and long-time translational dynamics vary non-monotonically with $\phi$ due to the competition between hydrodynamic and excluded-volume interactions.
Therefore, the suspension is most diffusive at a certain $\phi$, which is between 10\% and 20\% in our case.
This is in contrast to suspensions of self-propelling particles, where the translational diffusivity typically reduces with $\phi$ due to increased rotational diffusion at higher $\phi$.
Although the latter is also found in our system, the stronger hydrodynamic coupling between the translational and rotational motion herein changes the overall dynamics qualitatively.

The essential physical mechanism for the above observations lies in: 
(i) the coupling between the long-time and short-time dynamics, $D = U_{rms}^2 t_u/3$; 
(ii) the proportionality between the translational and rotational relaxation times, $t_u \approx t_c \sim \tau_r \equiv d_r^{-1}$; 
(iii) the scaling of the persistence length of the ballistic motion, $\ell_p \sim U_{rms}t_c \sim \phi^{-1/3}$; and 
(iv) the scaling of instantaneous translational speed, $U_{rms} \sim \phi^{\nu}$. 
From these relationships we only need to estimate the value of $\nu$ to infer the scalings of $D$ and $d_r$ with respect to $\phi$, at least in the dilute regime.
Such an estimation is furnished by a statistical analysis valid at large distances, assuming the suspension is homogeneous and isotropic (which we have verified), while also taking into account the local correlations in the microstructure.
In more concentrated suspensions, the last two scalings (iii and iv) change substantially because of crowding and more frequent collisions between particles.

Finally, we believe the present study can be extended in a number of ways to explore the dynamics of apolar active suspensions in more complex conditions.
One open question concerns the effect of particle shape in dense suspensions.
Although long-range nematic order cannot develop spontaneously in any apolar systems, non-spherical particles tend to have a stronger local alignment due to steric interactions at higher concentrations, and such alignment may affect the dynamics when the activities of nearby particles differ.
Another interesting question is how the dynamics would be altered by a time-dependent external shear.
It is known that non-Brownian particles undergo diffusive motion akin to Taylor dispersion under simple shear, but can also self-organize into non-diffusive ``absorbing states" in oscillatory shear \citep{Pine_Nature_2005, Corte_NatPhys_2008}.
Whether particles endowed with an internal activity can exhibit similar dynamics and, if so, what the ensuing rheology may be (c.f.~\citet{Ge2021, Ge2022}) remain to be studied.

\begin{acknowledgments}
This research is supported by the Swedish Research Council via an International Postdoc Grant, No.~2021-06669VR.
The authors wish to thank S.~Bagheri, L.~Brandt, S.~S.~Ray, H.~Diamant, J.-L. Thiffeault, and S.~Guo for useful discussions during the early phase of this work.
\end{acknowledgments}


\appendix

\section{Instantaneous and long-time dynamics in binary active suspensions}
\label{sec:fp}

In Sections \ref{sec:short} and \ref{sec:long} of the main text, we presented the effect of volume fraction ($\phi$) on the instantaneous and long-time dynamics, mainly focusing on purely active suspensions ($f_p =0$).
In the following, we briefly describe the results when incorporating passive particles into the suspensions ($f_p >0$).

\begin{figure*}
  \centering
  \includegraphics[height=7.5cm]{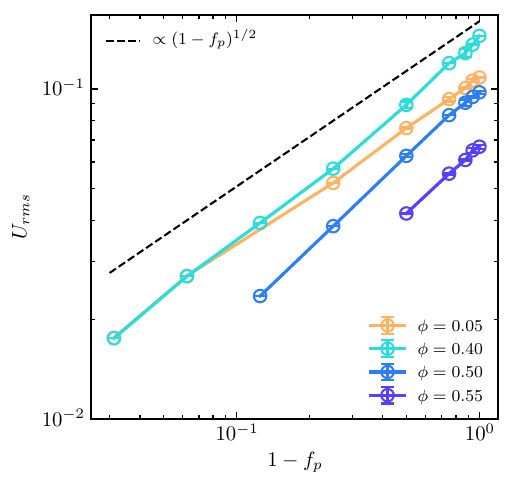}
  \includegraphics[height=7.5cm]{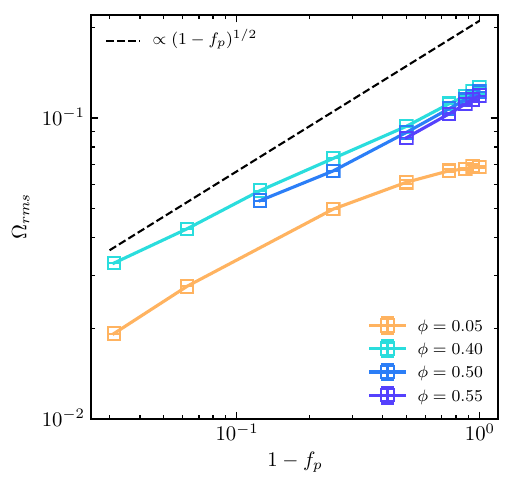}
  \begin{picture}(0,0)
  \end{picture} 
  \caption{Instantaneous speeds at different volume fractions ($\phi$) and ratios of passive particles ($f_p$).
               (a) Translational speed.
               (b) Rotational speed.}
  \label{fig:rms-fp}
\end{figure*}

Figure \ref{fig:rms-fp} shows the effect of $f_p$ on the instantaneous speeds at a few volume fractions.
As discussed in Section \ref{sec:short}, if the suspension is homogeneous and isotropic, we would expect the typical flow generated by neighboring active particles at a large distance $r$ to scale as $\sqrt{\phi_a}/r$, where $\phi_a \equiv \phi (1-f_p)$, leading to a flow proportional to $\sqrt{1-f_p}/r$ at a fixed $\phi$.
Therefore, if the active and passive particles are randomly mixed at all distances, we would expect both $U_{rms}$ and $\Omega_{rms}$ to scale as $(1-f_p)^{1/2}$.
This is observed in the case of $U_{rms}$ but not $\Omega_{rms}$, whose scaling exponent is less than $1/2$.
Since ${\bm \Omega}_p$ decays faster in $r$ than ${\bm U}_p$, the different scalings may be due to the local correlations in the microstructure, which have a larger effect on the final $\Omega_{rms}$ than $U_{rms}$ (same for the $\phi$ dependence).
Overall, it is clear that incorporating passive particles reduces the instantaneous particle speeds of the entire suspension.

\begin{figure*}
  \centering
  \includegraphics[height=7.5cm]{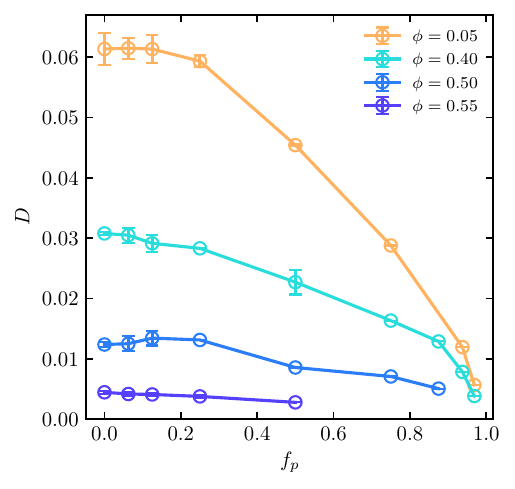}
  \includegraphics[height=7.5cm]{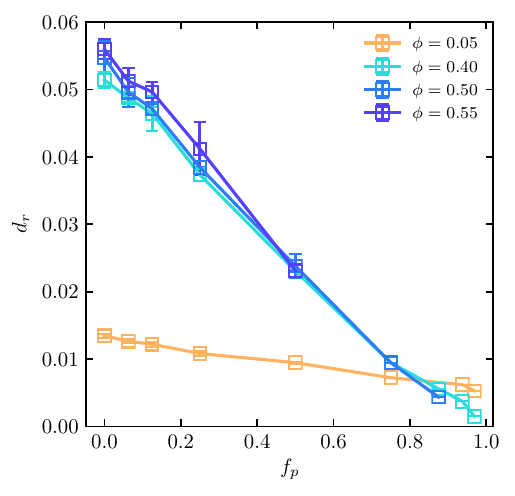}
  \begin{picture}(0,0)
  \end{picture} 
  \caption{Hydrodynamic diffusion at different volume fractions ($\phi$) and ratios of passive particles ($f_p$).
               (a) Translational diffusion coefficient.
               (b) Rotational diffusion coefficient.}
  \label{fig:D-fp}
\end{figure*}

Figure \ref{fig:D-fp} shows the long-time dynamics of binary suspensions at different volume fractions.
Similar to the instantaneous speeds, both $D$ and $d_r$ tend to reduce with $f_p$; however, the reduction of $D$ is nonlinear near $f_p=0$ and, at $\phi=0.5$, it even increases slightly with $f_p$.
To explain these behaviors, recall that the long-time and short-time dynamics are coupled, $D = U_{rms}^2 t_u/3$, and the rotational relaxation times are proportional, $t_u \approx t_c \sim \tau_r \equiv d_r^{-1}$.
Since $U_{rms} \approx U_{rms,0}(1-f_p)^{1/2}$ and $t_u \approx t_{u,0}(1 + \alpha f_p)$ for small $f_p$, where $\alpha>0$ (verified but not shown), we have $D \approx U_{rms,0}^2 (1-f_p) t_{u,0}(1 + \alpha f_p)/3 = D_0 (1-f_p) (1+\alpha f_p)$.
In these relations, $U_{rms,0}$, $t_{u,0}$, and $D_0$ correspond to $U_{rms}$, $t_{u}$, and $D$ at $f_p=0$, respectively.
Therefore, if $\alpha >1$ (\ie, $t_u$ increases rapidly with $f_p$), we may observe an enhanced translational diffusion when incorporating passive particles in an active suspension.

\section{Velocity distribution and nematic order}
\label{sec:nematic}

\begin{figure*}[t]
  \centering
  \includegraphics[width=\textwidth]{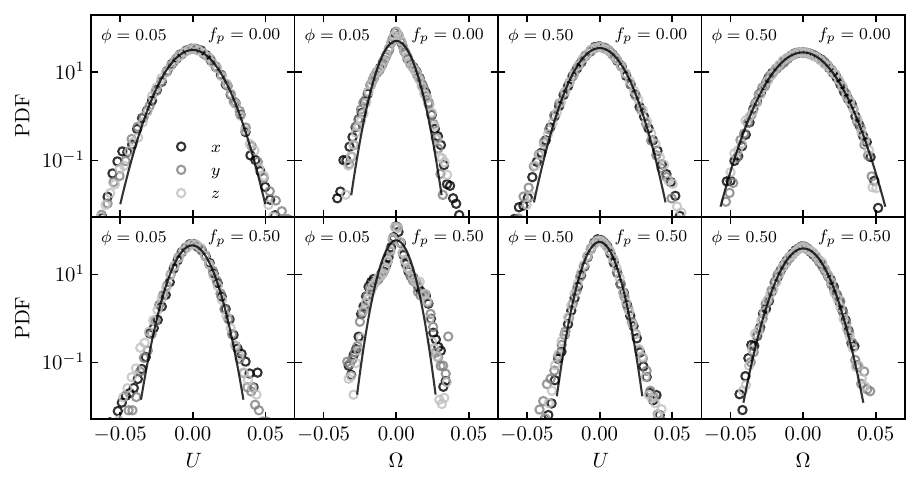}
  \caption{Distribution of particle velocities at different $\phi$ and $f_p$ for suspensions of pulling shakers and passive particles.
                Lines are Gaussian fits to the $x$-components.}
  \label{fig:vel-distr}
\end{figure*}

In Section \ref{sec:short} of the main text, we mentioned that the velocity distribution tends to be Gaussian if there are sufficiently many interacting active particles. 
Furthermore, the distribution of each velocity components should be the same as we have no reason to expect any difference (justified later). 
Figure \ref{fig:vel-distr} verifies these statements.
Specifically, we plot the three components of the instantaneous translational and rotational velocities, respectively, for $\phi=5\%$ or 50\% and $f_p=0$ or 0.5.
In each case, the different velocity components collapse onto the same distribution, which is always nearly Gaussian except for $\Omega$ when $\phi=5\%$.
In the latter cases ($f_p=0$ or 0.5), there is a higher concentration of slow particles, consistent with the bimodal speed distribution shown in Figure \ref{fig:speed-distr}.
These trends are generally observed in all cases that we have simulated.

\begin{figure*}
  \centering
  \includegraphics[height=7.6cm]{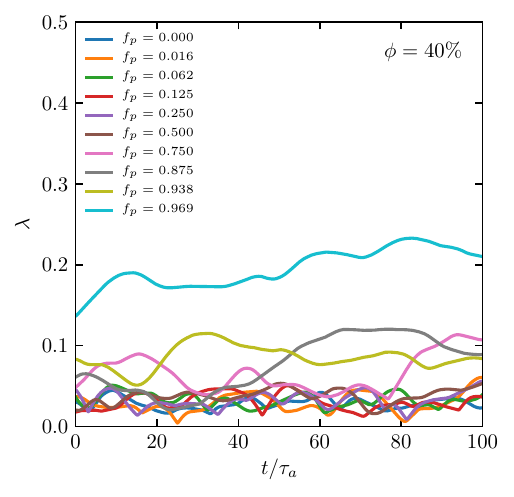}
  \includegraphics[height=7.5cm]{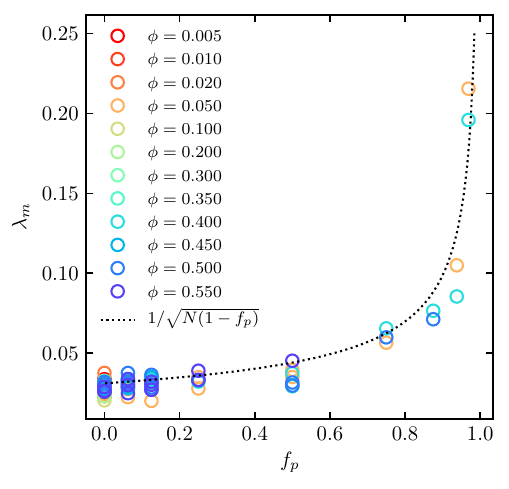}
  \begin{picture}(0,0)
    \put(-16,7){(a)} \put(-8.2,7){(b)}
  \end{picture} 
  \caption{Nematic order.
               (a) Time evolution of the nematic order parameter $\lambda$ for pullers and passive particles at $\phi=40\%$.
               (b) The mean nematic order $\lambda_m$ of all cases.}
  \label{fig:nematic}
\end{figure*}
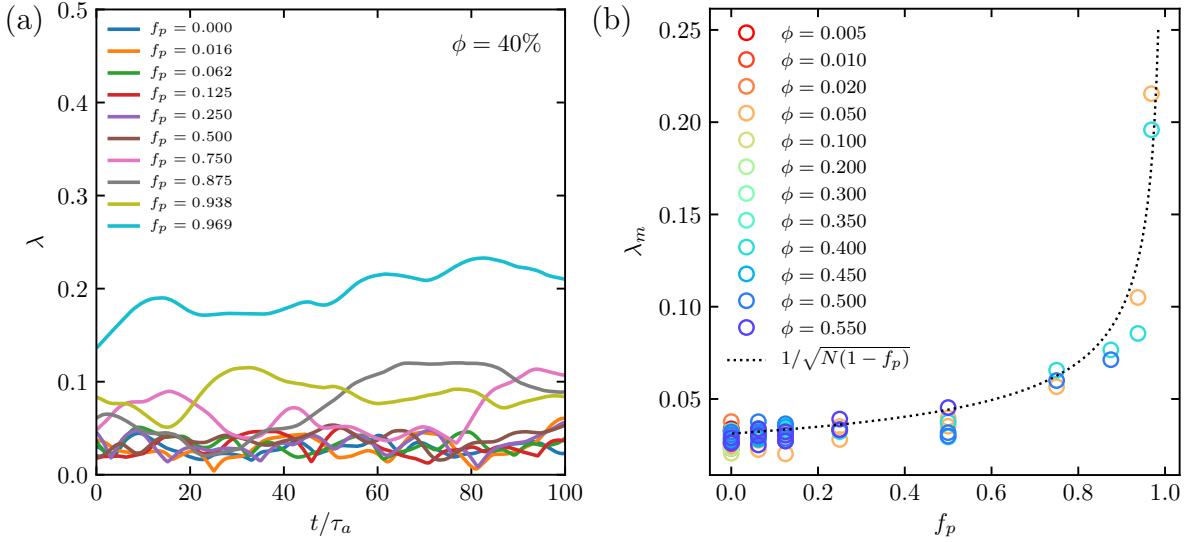

The reason we do not expect the distribution of the velocity components to differ is the absence of nematic order. 
For a binary suspension of shakers and passive particles, we can calculate a nematic tensor, $\bm Q$, given as
\begin{equation}
  Q_{ij}=\frac{3}{2} \bigg( \langle p_i p_j \rangle_{s} - \frac{1}{3}\delta_{ij} \bigg),
\end{equation}
where $\bm p$ is the orientation of the shaker, $\bm \delta$ the Kronecker delta, and $\langle \cdot \rangle_s$ denotes the average over all shakers.
By definition, $\bm Q$ is symmetric and traceless, and its largest eigenvalue, $\lambda$, measures the nematic order:
$\lambda=0$ if the suspension is isotropic; $\lambda=1$ if it is maximally nematic \citep{Doi2013}.
Figure \ref{fig:nematic} shows the time evolution of $\lambda$ for suspensions of pullers and passive particles at $\phi=40\%$ and the steady-state values ($\lambda_m$) at different $\phi$ and $f_p$.
Clearly, there is no nematic order in all cases as $\lambda_m$ is always within one standard error, $\sigma_\lambda \sim 1/\sqrt{N(1-f_p)}$.
Our data thus suggest that shakers remain isotropically oriented at all volume fractions and ratios of passive particles.
As a consequence, the individual velocity components must be identically distributed.


\bibliographystyle{apsrev4-2}
\bibliography{main}

\end{document}